\documentclass[aip,reprint,superscriptaddress,amsmath,amssymb]{revtex4-1}

\usepackage{graphicx}
\usepackage{dcolumn}
\usepackage{bm}
\usepackage{amsthm}
\usepackage[caption=false]{subfig}
\usepackage{algorithm}
\usepackage[noend]{algpseudocode}
\usepackage{float}
\usepackage[utf8]{inputenc}
\usepackage[T1]{fontenc}
\usepackage{mathptmx}
\usepackage[dvipsnames]{xcolor}
\usepackage{mathtools}

\DeclarePairedDelimiter\norm{\lVert}{\rVert}
\usepackage{theoremref}

\newtheorem{theorem}{Theorem}
\numberwithin{theorem}{section}

\usepackage{pifont}
\usepackage[toc,page]{appendix}

\usepackage{wrapfig}

\DeclareMathOperator*{\argmin}{arg\,min}

\newcommand{\doverlap}{\langle \phi_0 | e ^{-iHt_k} | \phi_0 \rangle}
\newcommand{\Kmax}{K_{\text{max}}}
\def\E{\mathbb{E}}
\def\I{\mathbb{I}}
\newcommand{\bxi}{\boldsymbol{\xi}}
\newcommand{\bmu}{\boldsymbol{\mu}}
\newcommand{\bw}{\mathbf{w}}
\newcommand{\bx}{\mathbf{x}}

\newcommand{\bz}{\mathbf{z}}
\newcommand{\bI}{\mathbf{I}}
\newcommand{\bC}{\mathbf{C}}
\newcommand{\bSigma}{\boldsymbol{\Sigma}}
\newcommand{\bGamma}{\boldsymbol{\Gamma}}
\newcommand{\MVN}{\mathsf{N}}
\newcommand{\CN}{\mathsf{CN}}
\DeclareMathOperator{\Erf}{Erf}

\newcommand{\rmd}{D}
\DeclarePairedDelimiter\floor{\lfloor}{\rfloor}

\makeatletter
\def\namedlabel#1#2{\begingroup
   \def\@currentlabel{#2}%
   \label{#1}\endgroup
}
\makeatother

\makeatletter
               {\list{}{\leftmargin=5pt 
                        \labelwidth\z@ \itemindent-\leftmargin
                        }}%
               {\endlist}
\makeatother

\makeatletter
\newcommand{\thickhline}{%
	\noalign {\ifnum 0=`}\fi \hrule height 1pt
	\futurelet \reserved@a \@xhline
}
\newcolumntype{"}{@{\hskip\tabcolsep\vrule width 1pt\hskip\tabcolsep}}

\makeatletter
\def\blfootnote{\gdef\@thefnmark{}\@footnotetext}
\makeatother

\usepackage{hyperref}
\hypersetup{colorlinks,linkcolor={blue},citecolor={olive},urlcolor={blue}}
\usepackage[capitalize,nameinlink]{cleveref}
\Crefname{figure}{Fig.}{Figs.}

\begin{document}


\title{From noisy observables to accurate ground‑state energies: a  quantum-classical signal subspace approach with denoising}
${}$
\vspace{0.5cm}
\author{Hardeep Bassi}\email{hbassi2@ucmerced.edu}\thanks{Work done as a visiting student researcher at Lawrence Berkeley National Laboratory}
\affiliation{Department of Applied Mathematics, University of California, Merced, CA 95343, USA}

\author{Yizhi Shen}\email{yizhis@lbl.gov}
\affiliation{
Applied Mathematics and Computational Research Division,
Lawrence Berkeley National Laboratory, Berkeley, CA 94720, USA}

\author{Harish S. Bhat}\email{hbhat@ucmerced.edu}
\affiliation{Department of Applied Mathematics, University of California, Merced, CA 95343, USA}

\author{Roel Van Beeumen}\email{rvanbeeumen@lbl.gov}
\affiliation{
Applied Mathematics and Computational Research Division,
Lawrence Berkeley National Laboratory, Berkeley, CA 94720, USA}


\begin{abstract}
We propose a hybrid quantum-classical algorithm for ground state energy (GSE) estimation that remains robust to highly noisy data and exhibits low sensitivity to hyperparameter tuning. Our approach---Fourier Denoising Observable Dynamic Mode Decomposition (FDODMD)---combines Fourier-based denoising thresholding to suppress spurious noise modes with observable dynamic mode decomposition (ODMD), a quantum-classical  signal subspace method. By applying ODMD to an ensemble of denoised time-domain trajectories, FDODMD reliably estimates the system's eigenfrequencies. We also provide an error analysis of FDODMD. Numerical experiments on molecular systems demonstrate that FDODMD achieves convergence in high-noise regimes inaccessible to baseline methods under a limited quantum computational budget, while accelerating spectral estimation in intermediate-noise regimes. Importantly, this performance gain is entirely classical, requiring no additional quantum overhead and significantly reducing overall quantum resource demands.

\end{abstract}
\maketitle

\section{Introduction}
\subsection{Overview}
Ground state estimation is a fundamental challenge across numerous branches of science and engineering, including chemistry, physics, and materials science. Understanding the ground state of a many-body quantum system provides critical insights into the system's properties. While classical diagonalization of the exact many-body Hamiltonian scales exponentially in the system size, quantum computing has the potential to solve such eigenvalue problems with theoretically guaranteed polynomial scaling~\cite{cao2019quantum, lee2023evaluating, reiher2017elucidating, lin2022lecture}. These  guarantees, however, rest on the core assumption of perfect fault-tolerance and error correction. As such capabilities remain out of reach for practical quantum hardware, there is a growing need to leverage current devices, despite their individual limitations. Hybrid quantum-classical approaches have been developed to exploit the capabilities of current noisy intermediate-scale quantum (NISQ) platforms. The idea of hybrid algorithms is to delegate to NISQ devices routines that best use their inherent strengths, for example representing and simulating many-body wave functions, and then to employ classical computers to post-process the results measured on the NISQ device~\cite{getelina2024quantum, park2024efficient, klymko2022real, ding2023even, shen2023estimating,mcclean2016theory,tilly2022variational, ding2023simultaneous, ding2024quantum, shen2023real}. 

One hybrid method for ground state estimation is  observable dynamic mode decomposition (ODMD)~\cite{shen2023estimating}.  ODMD assumes that one has already produced a time series of scalar observables via successive time evolution of a reference state.  The observables can be computed cheaply on quantum hardware, without computing full time-evolved states in an exponentially large Hilbert space.   Next, on classical hardware, one applies a variant of dynamic mode decomposition (DMD)~\cite{schmid2010dynamic, kutz2016dynamic} to post-process the time series of observables.  In the idealized, noise-free setting, the observables evolve as a linear combination  of complex exponentials, where each term corresponds to an eigenfrequency.  ODMD serves as a system identification and signal processing tool: when applied to a time series consisting of complex sinusoids, ODMD recovers the leading eigenfrequency component that contains the ground state energy (GSE).

Numerically, ODMD involves solving one least squares problem and one eigenvalue problem; neither of these steps are computationally intensive.  ODMD is exponentially convergent in the noise-free setting and resilient to perturbative noise~\cite{shen2023estimating}. As noise increases, this resilience weakens, as reflected in ODMD's reduced stability and reliance on hyperparameter tuning within a delicate, often unforgiving range.  Both theory and empirical findings motivate the search for a \emph{denoising} procedure that one can apply to the noisy time series of quantum observables to bring it closer to its ideal, noise-free counterpart.

The primary contribution of this paper is to augment ODMD with Fourier-based denoising procedures. In regimes where ODMD struggles to converge---especially under suboptimal hyperparameter choices---\emph{these denoising procedures substantially improve GSE estimation accuracy without requiring any extra quantum resources}.  One of our key ideas is to apply \emph{hard thresholding} to the observable time series' discrete Fourier transform (DFT).  Low-amplitude frequency components are interpreted as spurious noise modes and truncated away~\cite{holton2021digital, stephane1999wavelet}.  The other key idea is \emph{stacking}, which uses ODMD with multiple denoised realizations of the noisy time series.  Here each realization corresponds to denoising with a different threshold in Fourier domain.  As we show through multiple empirical tests, for noisy data where ODMD exhibits slow or no convergence, the combination of hard thresholding and stacking enables fast convergence to the true GSE.  As an aside, we also demonstrate that an enhancement called \emph{zero-padding}~\cite{holton2021digital} mitigates grid resolution issues that arise when estimating the GSE using Fourier-based methods.

\subsection{Related Work}
To our knowledge, this work presents the first use of Fourier-based denoising via frequency domain truncation in quantum algorithms for GSE estimation.  Fourier-based approaches, in the context of quantum signal processing, have been applied to various computational tasks~\cite{silva2022fourier, motlagh2024generalized, alexis2024quantum}. Recent work has incorporated Fourier-based  noise suppression via frequency domain windowing into a hybrid workflow to recover accurate single-particle spectra from noisy quantum correlator data~\cite{wang2025computing}. Wang \emph{et al.}\ use Hamming‑window filtering and cross‑spectral principal‑component analysis to re-weight, but not discard frequency components.  
In contrast, our method applies a hard threshold to remove low‑amplitude Fourier modes outright, producing denoised trajectories that serve as the signal subspace for subsequent DMD.

There exist numerous classical techniques~\cite{Stoica2005} that estimate the frequency set $\{ \omega_j \}$ of a time series consisting of a linear combination of complex exponentials $\{ \exp(i \omega_j t) \}$.  Classical Fourier-based estimation methods (such as periodograms) have a frequency resolution proportional to $1/K$, where $K$ is the length of the time series.  DMD is an example of a super-resolution method that goes beyond the classical limit; other examples, which share structural features with DMD, include MUltiple SIgnal Classification (MUSIC)~\cite{schmidt1986multiple,liao2016music}, Prony's method~\cite{prony1795essai, 
el1995new}, and Estimation of Signal Parameters via Rotational Invariant Techniques (ESPRIT)~\cite{roy1989esprit, li2020super,  ding2024esprit}.  In the idealized noise-free regime, these algorithms' computational complexity is proportional to the signal's sparsity in the frequency domain, not the number of samples $K$.  This ability to estimate spectra accurately from only a limited number of samples is a hallmark of super-resolution techniques.

A variety of hybrid quantum-classical algorithms~\cite{klymko2022real,shen2023real,ding2023even,shen2023estimating,ding2024quantum} have been recently developed and analyzed for eigenenergy estimation, offering alternatives that bypass stringent fault-tolerance requirements of standard quantum phase estimation (QPE)~\cite{kitaev1995quantum}.
In contrast to leading hybrid methods that solve generalized eigenvalue~\cite{klymko2022real,shen2023real} or non-linear optimization~\cite{ding2023even,ding2023simultaneous} problems to recover eigenenergies, ODMD casts
the task as a matrix least-squares problem. This formulation counteracts ill-conditioning often associated with nonorthogonal bases in generalized eigenvalue problems, while also sidestepping the computational overhead incurred by non-linear optimization.

We remark that some methods~\cite{ding2023even,ding2023simultaneous,ding2024quantum} further require that the initial state $|\phi_0\rangle$ must have dominant overlap ($p_0 > 0.5$) with the ground state for desirable convergence guarantees. This non-trivial requirement may demand substantial quantum resources at the reference state preparation stage~\cite{lin2020near}. By comparison, ODMD is provably convergent without such dominance assumptions.  In addition to eigenenergy estimation, ODMD canonically incorporates eigenstate recovery, noise source estimation, super-resolved spectral analysis, and better conditioning via a least-squares framework; to our knowledge, no other hybrid quantum-classical or Fourier-based method delivers all four of these advantages simultaneously. 
 
Overall, this work establishes a robust hybrid algorithm that leverages frequency domain denoising with a quantum signal subspace approach to improve many-body ground state energy estimation from noisy quantum data. Throughout this paper, we focus our numerical demonstrations on molecular datasets, which underpin electronic structure calculations in quantum chemistry and pose unique challenges for classical algorithms. Unlike familiar spin Hamiltonians where interactions are local and amenable to classical compression techniques such as tensor networks, molecular Hamiltonians involve long-range Coulomb interactions and system-specific orbitals that extend polynomially with the total count of electrons. These features can lead to highly entangled and/or strongly correlated many-body states, implying a steep computational cost and making molecular systems compelling targets for hybrid quantum algorithms~\cite{larsson2022chromium,mccaskey2019quantum, larsson2022matrix, baiardi2019large}.

For the specific molecules examined in this work, we focus on GSE estimation in the low ground state overlap, intermediate-to-high noise regime. Combining Fourier denoising with ODMD is especially effective here, enabling desired accuracy and more optimal use of NISQ platforms with a significantly reduced computational budget.

\subsection{Organization}
In \cref{sec:bg}, we review necessary background including the measurement of quantum observables via the Hadamard test, Fourier representation of time series, and basic noise models perturbing the observables. In \cref{sec:methods}, we describe our methodology. We first overview the ODMD approach for solving the ground state problem. Next, we introduce our novel Fourier denoising ODMD (FDODMD) approach, which integrates frequency domain thresholding with signal stacking. We also describe a time domain zero-padding procedure to enhance Fourier-based ground state energy estimation.  In \cref{sec:data}, we detail how we generate data and how we test the convergence of algorithms.  We discuss in \cref{sec:depolarizing} a motivating example that involves data with both shot noise and depolarizing error.  In \cref{sec:results}, we compare FDODMD with baseline ODMD and classical (including zero-padded) Fourier-based spectral estimation for the chromium dimer.  In \cref{sec:formal_analysis}, we derive error bounds that help to explain  the performance improvements of FDODMD.  \Cref{sec:conclusion} summarizes our findings and discusses their broader implications, along with directions for future work.  In \cref{app:optimal_shot_allocation,app:scaling,app:supplementary_figs,app:formal_proofs}, we discuss optimal shot allocation, Hamiltonian scaling, results for LiH, and formal proofs of bounds, respectively.

\section{Background}\label{sec:bg}
\subsection{Acquiring data from a quantum circuit}
Consider the expectation values
\begin{equation}\label{eqn:unitaryoverlap}
    \langle U \rangle = \langle \phi_0 | U | \phi_0 \rangle, 
\end{equation}
where $|\phi_0\rangle$ is a fixed reference quantum state and $U$ is a unitary operator. Given a many-body Hamiltonian $H$, we can efficiently simulate the one-parameter family of time evolution operators $U(t) = e^{-iHt}$ on quantum hardware due to its native unitarity. These real-time expectation values $\langle U(t_{k}) \rangle$ evaluated at discrete times $t_k$ form the basis of several recently proposed hybrid algorithms for energy estimation~\cite{shen2023estimating,shen2024efficient,ding2023even,klymko2022real}.

\begin{figure*}[t]
    {\centering%
    \includegraphics[width=0.95\linewidth,clip,trim=125 50 160 20]{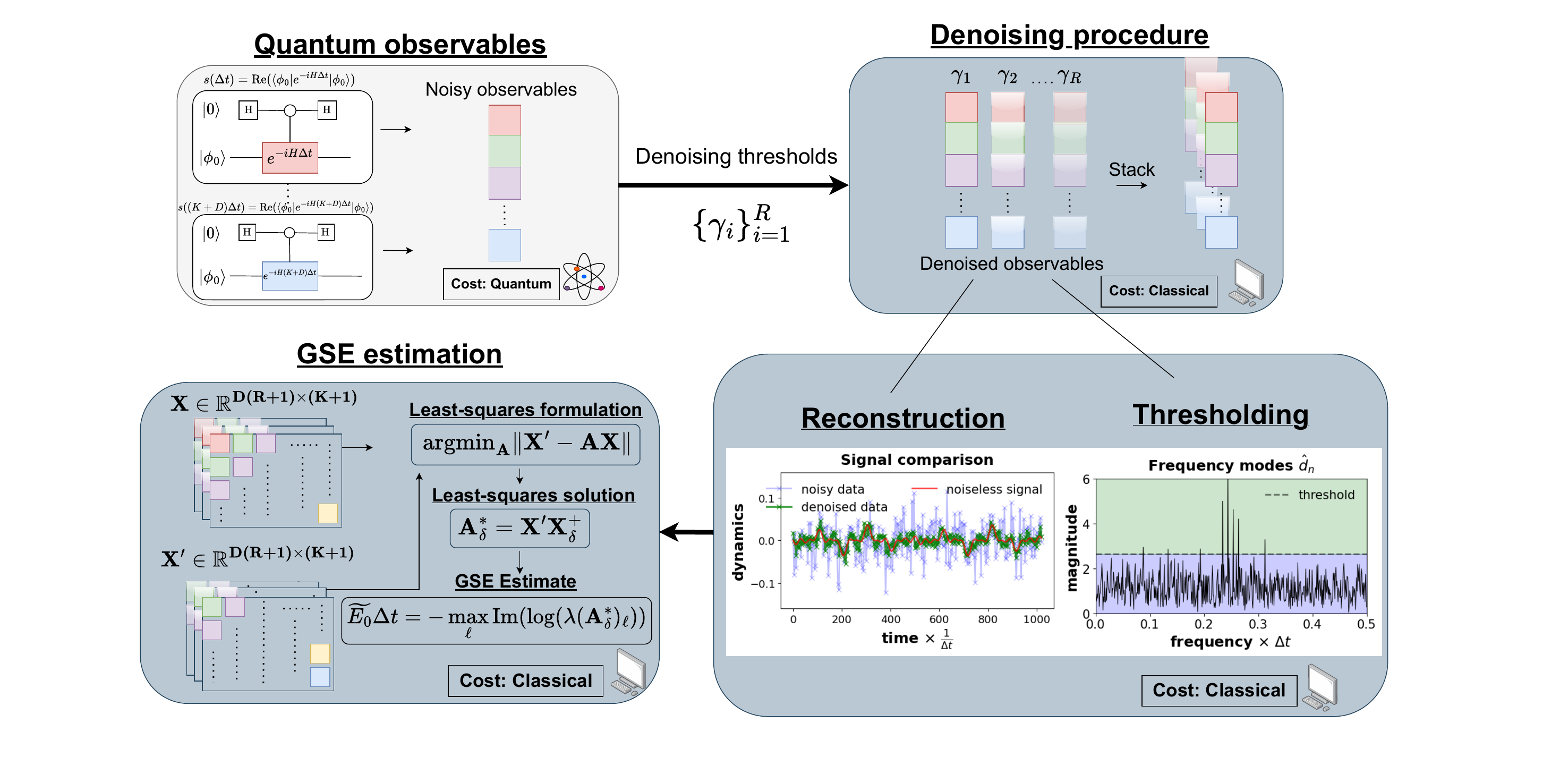}}
    \caption{Total workflow of FDODMD as outlined in \cref{sec:methods}. \textbf{(Top left)} The Hadamard test is used to obtain noisy measurements of the data (\ref{eqn:doverlapnoisy}). \textbf{(Top right)} The denoising procedure is then applied by taking $R$ different truncation factors in (\ref{eqn:dift}). The data is then stacked together according to (\ref{eqn:signalstacked}). \textbf{(Bottom right)} Here we illustrate the thresholding and denoised reconstruction procedures described in \cref{sect:denoising}. \textbf{(Bottom left)}  The stacked data is used to estimate the GSE, as described in \cref{sect:modmd}. }
    \label{fig:odmd}
\end{figure*}

To access the overlap (\ref{eqn:unitaryoverlap}) on a quantum device, we employ the Hadamard test~\cite{lin2022lecture, cleve1998quantum}, a standard subroutine for measuring the expectation value of a general unitary operator. Since (\ref{eqn:unitaryoverlap}) is complex-valued, two Hadamard test circuits are executed, one for the real part and one for the imaginary part. The Hadamard test is visualized in the upper left panel (light gray) of \cref{fig:odmd}. The circuits use a single ancilla and, for each shot, 
yield a binary outcome corresponding to the measurement of the ancilla in the $|0\rangle$ or $|1\rangle$ basis state, respectively. Averaging over many shots generates an unbiased estimate for the desired overlap. For our purposes, we run the classically simulated Hadamard test to obtain time series of quantum observables.


The approaches introduced in this paper require as input a molecular Hamiltonian $H$ and a reference state $|\phi_0\rangle$ with overlap probability $p_0 = |\langle \psi_0 | \phi_0\rangle |^2$, where $|\psi_0 \rangle$ is the true ground state.  Suppose we have data on an equispaced temporal grid with $t_k = k \Delta t$ for $0 \leq k \leq K$.  Given the above, we measure the real and/or imaginary parts of the \emph{signal}:
\begin{equation}\label{eqn:doverlap}
    s(t_k) = \langle U(t_k) \rangle = \langle \phi_0 | e^{-iHk\Delta t} | \phi_0 \rangle.
\end{equation}
In practice, however, measurements performed on current quantum hardware are subject to various sources of error~\cite{de2021materials, resch2021benchmarking, johnstun2021understanding, georgopoulos2021modeling}. Two of the most prevalent sources are gate error,  which results from imperfect quantum operations, and shot noise, which reflects the statistical fluctuations of a finite sampling of shots. Since precisely modeling gate error is highly device-specific and technically challenging~\cite{georgopoulos2021modeling, resch2021benchmarking}, we focus primarily on shot noise and examine a simplified global model of gate error.

As a result, when implementing the Hadamard test on NISQ hardware, we measure a noisy version of the real-time overlap (\ref{eqn:doverlap}), which we write schematically as the \emph{data}:
\begin{equation}
\label{eqn:doverlapnoisy}
d(t_k) = s(t_k) + \xi(t_k),
\end{equation}
where $\xi(t_k)$ represents the noise. We emphasize that the input data for all our methods are either the noisy observables (\ref{eqn:doverlapnoisy}), or a denoised version of them as described in \cref{sec:methods}.  

\subsection{Noise models}
\label{sect:noisemodels}
Here we discuss how to model the quantum noise $\xi(t_k)$ when constructing (\ref{eqn:doverlapnoisy}).  In principle, $s(t_k)=\doverlap$ is a complex-valued scalar.  For simplicity, we focus on the real part; the imaginary part can be treated analogously.

Since the Hadamard test yields outcomes $\pm 1$, the readout of shot $i$ at time $t_k$ is a Bernoulli random variable $b_i(t_k)$ that takes values $+1$ and $-1$ with respective probabilities $\pi_{t_k}$ and $1-\pi_{t_k}$. To ensure that the Bernoulli estimator is unbiased, we require $\mathbb{E} [b_i(t_k)] = \Re s(t_k)$, which implies $\pi_{t_k} = (1 + \Re s(t_k))/2$.

We assume that the random variables $\{b_i(t_k)\}$ are independent and identically distributed.  At time $t_k$, the noisy observable $\Re d(t_k)$ is the empirical mean over a total of $n_{t_k}$ shots:
\begin{equation}\label{eqn:noisyoverlap2}
   \Re d(t_k) = \frac{1}{n_{t_k}} \sum_{i = 1}^{n_{t_k}} b_i(t_k).
\end{equation}
One can now check that $\E[\Re d(t_k)] = \Re s(t_k)$ and that
\begin{equation}\label{eqn:varshots}
\text{Var}(\Re d(t_k)) = 4 \pi_{t_k}(1-\pi_{t_k}) / {n_{t_k}},
\end{equation}
which vanishes as $n_{t_k} \to \infty$.  This confirms that in the infinite-shot limit, $\Re d(t_k)$ converges to $\Re s(t_k)$, as desired. 

To emulate shot noise in our simulations, we compute the exact value of (\ref{eqn:doverlap}) and then add zero-mean Gaussian noise,
\begin{equation}
\label{eqn:noisyoverlap1}
\Re d(t_k) = \Re \doverlap + \xi_k, \quad \xi_k \sim \mathcal{N}(0, \epsilon),
\end{equation}
where the standard deviation $\epsilon$ can be chosen to match $\sqrt{\text{Var}(\Re d(t_k))}$ from (\ref{eqn:varshots}). By the central limit theorem, the Gaussian noise model (\ref{eqn:noisyoverlap1}) provides a close approximation to the scaled binomial model (\ref{eqn:noisyoverlap2}). 
For this reason, we adopt the simple Gaussian model of (\ref{eqn:noisyoverlap1}) in our numerical experiments. \Cref{app:optimal_shot_allocation} discusses how to allocate shots across time, and finds that assigning an equal number of shots to each time step performs nearly optimally.

\subsection{Fourier data interpretation}
Upon expanding (\ref{eqn:doverlap}) in the eigenbasis of the Hamiltonian $H$, the observables considered in this work are linear combinations of complex exponentials, 
\begin{equation}\label{eqn:sumcomexp}
s(t_k) = \sum_{n=0}^{N-1}p_n e^{-iE_n t_k}, \quad t_k = k \Delta t,
\end{equation}
where $E_n$ denote the Hamiltonian eigenvalues, and $p_n$ denote the squared overlaps between the Hamiltonian eigenstates and reference state, i.e. $p_n = |\langle \phi_0 | \psi_n \rangle|^2$ with $H | \psi_n \rangle = E_n | \psi_n \rangle$. To analyze a sequence of observables $(s(t_0), s(t_1), \ldots, s(t_{K}))$ in the frequency domain, we can compute its discrete Fourier transform (DFT):
\begin{equation}
\label{eqn:dftsumcomexp}
\hat{s}(f_m) = \sum_{k=0}^{K}s(t_k)e^{-i f_m t_k}, \quad  f_m = \frac{2\pi m}{(K+1) \Delta t},
\end{equation}
where $f_m$ for $0 \leq m \leq K$ are  discrete frequency bins. The spacing between neighboring bins defines the associated frequency resolution $\Delta f = 2\pi ((K+1) \Delta t)^{-1}$. 

Under typical discretization schemes for the sequence length $K+1$ and time step $\Delta t$, the ground state energy $E_0$ \textit{does not} align exactly with any of the frequency bins $f_m$. Accordingly, its contribution is spread across multiple frequency bins. This spectral leakage can be exacerbated by the presence of noise.   Using the above, we can write the DFT of the noisy data (\ref{eqn:doverlapnoisy}), which we will use below in \cref{sect:denoising}:
\begin{equation}
\label{eqn:noisydft}
     \hat{d}(f_m) = \sum_{n=0}^{N-1}p_n\sum_{k=0}^{K}e^{-i(E_n + f_m)t_k } + \sum_{k=0}^{K}\xi(t_k)e^{-i  f_m t_k}.
\end{equation}

\section{Methods}\label{sec:methods}
\subsection{Observable dynamic mode decomposition (ODMD)}
\label{subsec:ODMD}

Observable dynamic mode decomposition (ODMD)~\cite{shen2023estimating} is a specialized adaptation of dynamic mode decomposition (DMD) for energy estimation, formulated in the space of quantum observables. DMD is a classical data-driven technique that extracts dominant modes of a dynamical system from real-time measurements and has demonstrated success across various applications~\cite{schmid2010dynamic,kutz2016dynamic, hawinkel2015time, grosek2014dynamic,brunton2016compressed,Yin2023}. 

DMD assumes access to snapshots of the full state of the dynamical system.
Suppose we observe a dynamical system with states $\mathbf{x}(t) \in \mathbb{R}^N$ at discrete times $t= t_0, t_1, \ldots, t_K$.  Given the time snapshots $\{\mathbf{x}(t_k) \}_{k=0}^K$, we arrange them into a pair of time-shifted data matrices,
\begin{align}
\label{eqn:xmat}
X &= \begin{bmatrix}
\mathbf{x}(t_0) & \mathbf{x}(t_1) & \cdots & \mathbf{x}(t_{K-1})
\end{bmatrix}, \\
\label{eqn:xprimemat}
X' &= \begin{bmatrix}
\mathbf{x}(t_1) & \mathbf{x}(t_2) & \cdots & \mathbf{x}(t_{K})
\end{bmatrix}.
\end{align}
with $X,X' \in \mathbb{R}^{N \times K}$.  We see that $X'$ is $X$ evolved forward by one time step. The forward propagator of the dynamics is then estimated via the solution of the least-squares problem
\begin{equation}\label{eqn:Asol}
A^\ast = \argmin_A \norm{X' - AX}_F.
\end{equation}
The formal solution to (\ref{eqn:Asol}) is given by $A^\ast = X'X^+$, with $X^+$ being the Moore-Penrose pseudoinverse of the data matrix $X$. In practice, we compute the pseudoinverse via a truncated singular value decomposition (SVD) using a threshold $\delta \sigma_{\text{max}}$, where $\sigma_{\text{max}}$ is the maximal singular value of $X$ and $\delta \in (0,1)$. This yields the regularized DMD solution $A^\ast_{\delta} = X' X_{\delta}^+$, where $X_{\delta}^+$ represents the truncated pseudoinverse of $X$.

For quantum systems, however, snapshots of the full state would require access to the full many-body states $|\phi(t)\rangle$ at different times; recall that such states live in an intractably high-dimensional Hilbert space. Thus, to adapt DMD to this setting, ODMD considers scalar observables of the state at different times rather than the full state snapshots. Specifically, given a Hamiltonian $H$ with time-evolution operator $e^{-iHt}$, we seek observables consisting of complex-valued overlaps $s(t_k)$ from (\ref{eqn:doverlap}). These can be estimated with the simulated Hadamard test mentioned in \cref{sec:bg}, which measures the real and imaginary parts of the overlap and yields a noisy time series of quantum observables $d(t_k)$ as in (\ref{eqn:doverlapnoisy}).

Using this noisy observable time series, we now construct time-shifted matrices by specifying a delay parameter $\rmd \ll N$. The data matrices (\ref{eqn:xmat}) and (\ref{eqn:xprimemat}) then take the form
\begin{align}
\label{eqn:newxmat}
X &= \begin{bmatrix}
d(t_0) & d(t_1) & \cdots & d(t_{K}) \\
d(t_1) & d(t_2) & \cdots & d(t_{K+1}) \\
\vdots & \vdots &  & \vdots \\
d(t_{\rmd-1}) & d(t_{\rmd}) & \cdots & d(t_{K+\rmd-1})
\end{bmatrix}, \\
\label{eqn:newxprimemat}
X' &= \begin{bmatrix}
d(t_1) & d(t_2) & \cdots & d(t_{K + 1}) \\
d(t_2) & d(t_3) & \cdots & d(t_{K + 2}) \\
\vdots & \vdots &  & \vdots \\
d(t_{\rmd}) & d(t_{\rmd+1}) & \cdots & d(t_{K+\rmd})
\end{bmatrix},
\end{align}
where $X,X' \in \mathbb{C}^{\rmd \times (K+1)}$ admit a Hankel structure. To form these matrices, we need data collected at times $t_k$ with $k=0, \ldots, K+D$; thus when using ODMD, all expressions of $K$ from \cref{sec:bg} should be thought of as $K+D$. Comparing (\ref{eqn:newxmat}-\ref{eqn:newxprimemat}) to (\ref{eqn:xmat}-\ref{eqn:xprimemat}), we notice that each column of our new data matrices $[d(t_j), d(t_{j+1}), \ldots, d(t_{j+\rmd-1})]^T$ plays the role of the full state, analogous to $\mathbf{x}(t_j)$ in the classic formulation. This technique is a form of time-delay embedding, which has been studied extensively in the literature~\cite{Kamb2020}. Even though the delay parameter $\rmd$ does not satisfy the conditions required for full state reconstruction~\cite{takens1981detecting,Kantz_Schreiber_2003}, it is sufficient to capture relevant spectral features.  

Notably, Koopman operator theory~\cite{Arbabi2017,Brunton2022} reveals that eigenvalues of $A^\ast \in \mathbb{C}^{\rmd \times \rmd}$, obtained by solving (\ref{eqn:Asol}) using the Hankel matrices (\ref{eqn:newxmat}) and (\ref{eqn:newxprimemat}), must closely approximate those of $H$. Note that $A^*$ serves as a proxy for the time evolution operator $e^{-iH \Delta t}$ whose eigenvalues are $e^{-iE_n \Delta t}$; since we approximate $A^\ast$ by the truncated SVD solution $A^\ast_{\delta} = X' X_{\delta}^+$, the ODMD estimate of the GSE is
\begin{equation}
\label{eqn:gseest}
    \tilde{E}_0 = -\max_{1\leq \ell \leq \rmd} \frac{\Im \log(\lambda(A^\ast_{\delta})_\ell)}{\Delta t},
\end{equation}
where $\lambda(A^*)_\ell$ denotes the $\ell$-th eigenvalue of $A^*$. The procedure for ODMD energy estimation is shown in the bottom left panel of \cref{fig:odmd}, where $R=0$.

\subsection{Frequency domain denoising}
\label{sect:denoising}

Fourier denoising ODMD (FDODMD) centers on two ideas, the first of which is frequency domain denoising.  The denoising itself is conceptually simple: given the data (\ref{eqn:doverlapnoisy}), we compute its DFT (\ref{eqn:noisydft}) and truncate.  Specifically, for a Fourier amplitude threshold $\tau_r > 0$, we truncate modes $f_m$ for which $|\hat{d}(f_m)| < \tau_r$. We heuristically choose $\tau_r =\gamma \cdot \text{median}(\hat{d}(f_m))$, where $\gamma > 0$ is a scalar. This choice is statistically meaningful and easily computable; setting $\gamma = 1$ provides an intuitive cutoff under reasonable assumptions on the signal-to-noise ratio. After truncating the DFT via
\begin{equation}\label{eqn:tdft}
\hat{r}(f_m) =
\begin{cases}
\hat{d}(f_m) & |\hat{d}(f_m)| \geq \gamma \cdot \text{median}(\hat{d}(f_m)) \\
0 & \text{otherwise},
\end{cases}
\end{equation}
we apply the inverse discrete Fourier transform (IDFT) to recover the denoised signal in the time domain: for $k=0, \ldots, K+D$,
\begin{equation}\label{eqn:dift}
r(t_k) = \frac{1}{K+D} \sum_{m \in \text{kept modes}} \hat{r}(f_m) e^{i f_m t_k}.
\end{equation} 
This will be referred to as the \textit{reconstruction} or \textit{denoised data}.  The above procedure maps one time series of quantum observables $\{ d(t_k) \}$ to one noise-suppressed realization $\{ r(t_k) \}$.  If we denoise the time series of observables multiple times---each time choosing a different threshold $\gamma$---we will obtain multiple noise-suppressed realizations.

\subsection{Signal stacking and multi-observable ODMD (MODMD)}
\label{sect:modmd}
The second idea underpinning FDODMD is signal stacking, which uses multiple denoised realizations with ODMD to estimate the GSE. For a collection of Fourier amplitude thresholds $\{\gamma_i\}_{i=1}^{R}$, we first generate the ensemble of denoised realizations $\{r_i(t_k)\}_{i=1}^{R}$, and then stack them alongside the original noisy signal to form the following extended multi-observable vector:
\begin{equation}\label{eqn:signalstacked}
\vec{d}(t_k) = 
\begin{bmatrix}
d(t_k) \\
r_1(t_k) \\
r_2(t_k) \\
\vdots \\
r_R(t_k)
\end{bmatrix}.
\end{equation}
We modify the shifted Hankel data matrices (\ref{eqn:newxmat}) and (\ref{eqn:newxprimemat}) by replacing each scalar entry $d(t_k)$ with the vectorization $\vec{d}(t_k)$ defined by (\ref{eqn:signalstacked}). This yields a pair of shifted block Hankel matrices $\mathbf{X}, \mathbf{X}' \in \mathbb{R}^{\rmd (R + 1) \times (K + 1)}$ that we use to formulate $\mathbf{A}^\ast = \argmin_{\mathbf{A}} \norm{\mathbf{X}' - \mathbf{A} \mathbf{X}}_F$, a least squares problem.  We compute the approximate solution $\mathbf{A}_\delta^\ast = (\mathbf{X}'\mathbf{X}_\delta^+) \in R^{\rmd(R + 1) \times \rmd(R + 1)}$ via the same truncated SVD scheme as in ODMD; our GSE estimate is (\ref{eqn:gseest}) with $\mathbf{A}^\ast_{\delta}$ in place of $A^\ast_{\delta}$.  The above constitutes MODMD, a multi-observable generalization of ODMD.

\Cref{fig:odmd} provides a schematic overview of the overall FDODMD denoising procedure.  Starting with quantum observable data in the upper-left panel, we proceed to the upper-right panel by denoising with $R$ different thresholds $\{ \gamma_i \}_{i=1}^{R}$.  This yields $R$ different denoised observable realizations, which can then be stacked.  For the purposes of following the FDODMD procedure itself, one can skip from the upper-right panel to the lower-left panel.  The stacked denoised realizations are used to assemble the block Hankel matrices $\mathbf{X}$ and $\mathbf{X}'$, which are then used to solve for $\mathbf{A}_\delta^\ast$ and produce the FDODMD estimate of the GSE.

In the lower-right panel of \cref{fig:odmd}, we illustrate both reconstruction and thresholding.  On the thresholding side, for one time series of quantum observables $\{ d(t_k) \}$, we plot the magnitude of the DFT $| \hat{d}(f_m) |$ as a function of frequency.  When we apply thresholding as in (\ref{eqn:dift}), sub-threshold modes are truncated to zero; super-threshold modes are kept without modification.  Not all discarded frequency components correspond purely to noise, implying a tradeoff between reconstruction fidelity and noise suppression that we analyze in \cref{sec:formal_analysis}.  On the reconstruction side, we plot three time series: the \emph{noiseless signal} $\{s(t_k)\}$, the \emph{noisy data} $\{d(t_k)\}$, and the \emph{denoised data} $\{r(t_k)\}$.  Note that the denoised data is closer to the noiseless signal than it is to the noisy data, showing the utility of our simple DFT-based denoising strategy.

\begin{figure}[t!]
    \centering
    \includegraphics[width=0.87\linewidth]{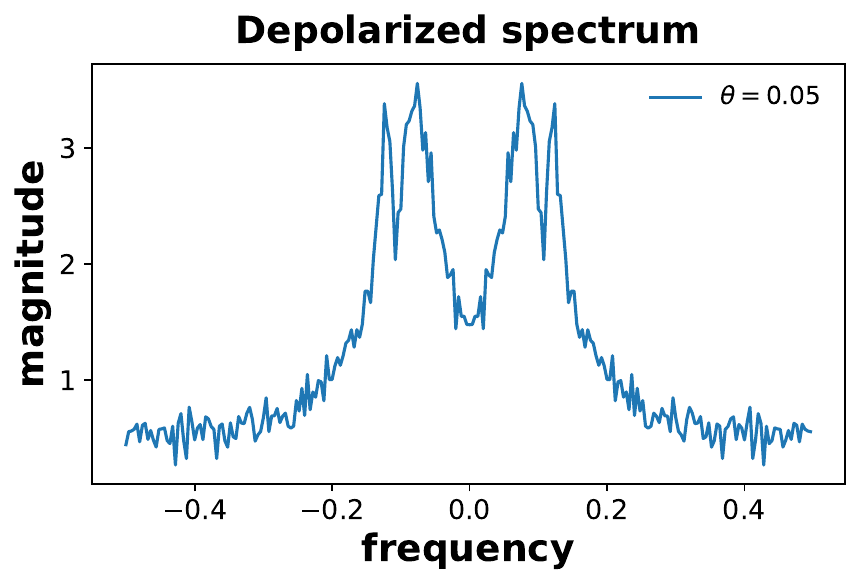}
    \caption{ In the presence of depolarizing error and shot noise, sharp peaks in the frequency domain are obfuscated. We plot the magnitude of the DFT of the noisy time series generated from (\ref{eqn:depolarizing}) with $(\theta, \Delta t, \Kmax, p_0) = (0.05, 1.0, 300, 0.2)$ and $\xi(t)\sim \mathcal{N}(0, \epsilon=0.01)$.}
    \label{fig:depolarized_spectrum}
\end{figure}

\begin{figure*}[t!]
\centering
\subfloat[GSE estimate]
{\includegraphics[width=0.48\textwidth]{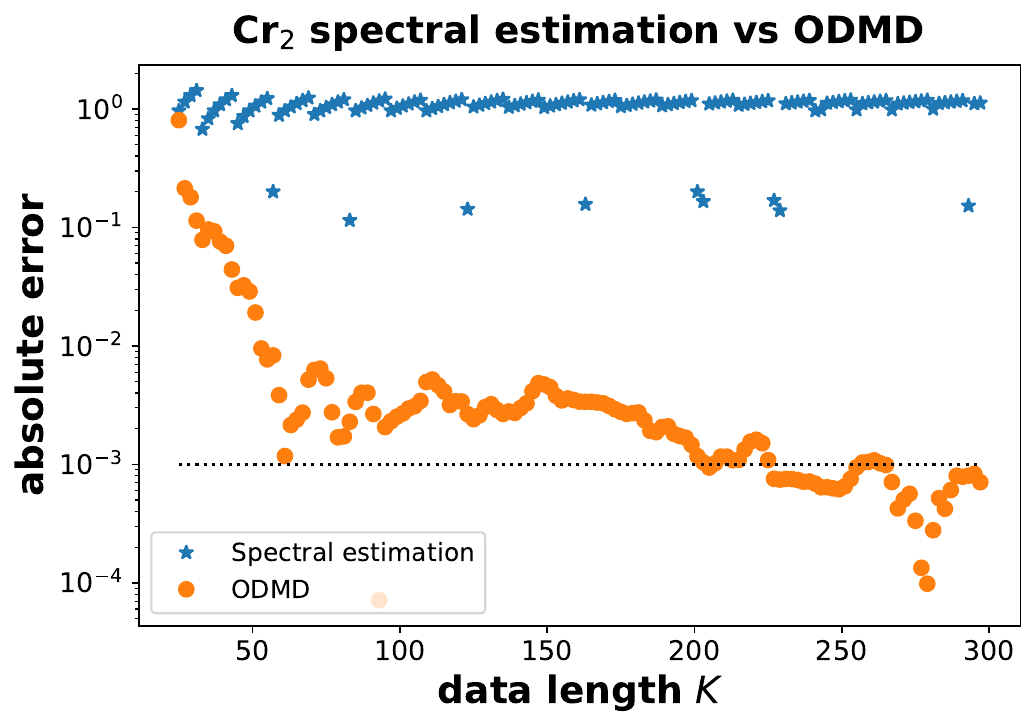}}
\subfloat[Depolarizing parameter estimate and dynamics reconstruction]
{\includegraphics[width=0.48\textwidth]{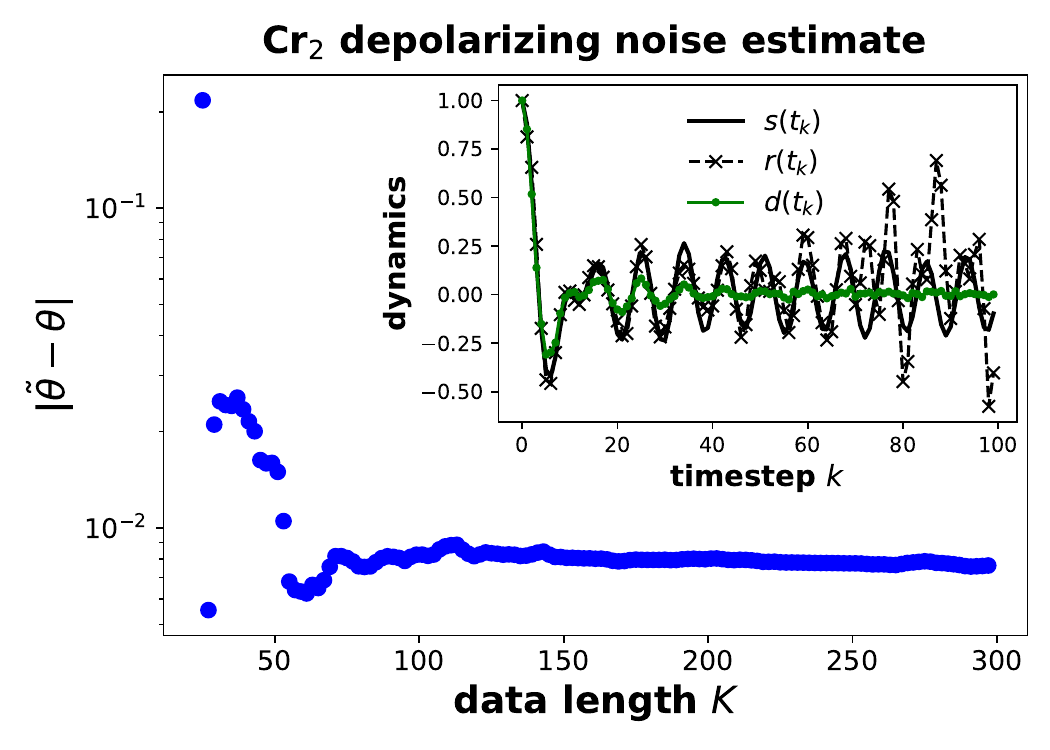}}
    \caption{ ODMD displays advantages in estimation and denoising for Cr$_2$ under depolarizing error and shot noise. 
 See the main text in \cref{sec:depolarizing} for details on data generation. To compute absolute errors as a function of data length $K$, we follow the testing protocol described in \cref{sec:data}. 
\textbf{(Left)} Convergence behavior of basic DFT-based spectral estimation (stars) and ODMD (circles). The dashed black line indicates chemical accuracy (absolute error $< 10^{-3}$.) ODMD converges quickly to chemical accuracy, in stark contrast to spectral estimation. \textbf{(Right)} Convergence of the estimate $\tilde{\theta}$ of the global depolarizing error parameter $\theta$ using ODMD. Inset in the upper right is the noisy data $\{d(t_k)\}$, the underlying signal $\{s(t_k)\}$, and the ODMD-reconstructed dynamics $\{r(t_k)\}$.  See \cref{sec:depolarizing} for details on how the reconstruction is computed. We see favorable agreement between the reconstruction and the noiseless signal, with deviations at long times due to imperfect denoising.}
    \label{fig:depolarizing_convergence}
\end{figure*} 

\subsection{Zero-padding in time domain}
As an alternative to the signal subspace method, we can enhance the performance of DFT-based estimation through a simple modification. In order to increase the frequency resolution of DFT \textit{without} altering the original signal content, we employ zero-padding, a useful technique in signal processing~\cite{holton2021digital}. Specifically, we augment our noisy time series by appending $M$ zeros, 
\begin{equation}\label{eqn:zeropadded}
d(t_k) =
\begin{cases}
d(t_k), & 0 \leq k \leq K+D, \\
0, & K+D < k \leq K+D + M.
\end{cases}
\end{equation}
This can be understood as modulating a time series of length $K+D+M+1$ with a rectangular window of length $K+D+1$. The DFT of this zero-padded time series becomes,
\begin{equation}\label{eqn:zeropaddeddft}
\hat{d}(f'_m) = \sum_{k=0}^{K + D + M}d(t_k)e^{-i f'_m t_k},
\end{equation}
with refined frequency bins $f'_m = 2 \pi m ((K + D + M + 1)\Delta t)^{-1}$ for $m = 0, \ldots, K + M + D$. We estimate the GSE by identifying the $f'_m$ that maximizes $|\hat{d}(f'_m)|$. This refinement improves  frequency resolution at no additional quantum cost, which potentially enables more accurate energy estimation.

Here we do not have the liberty of decreasing our time step $\Delta t \to 0$ to exactly resolve the GSE. The limitation arises for two reasons. First, reducing $\Delta t$ would require a proportionally larger number of quantum queries to reach a given evolution time, which is infeasible under limited quantum resources. Second, ODMD and FDODMD rely on capturing dynamical features over time, necessitating a sufficiently large $\Delta t$. This is in contrast to the smaller $\Delta t$ typically favored for DFT-based frequency estimation. To ensure a fair comparison, we thus keep $\Delta t$ fixed across methods, including zero-padded DFT.

\section{Data}
\label{sec:data}
We use classical methods and hardware to simulate noisy quantum observables.  In \cref{sec:depolarizing,sec:results}, we present results for the chromium dimer ($\text{Cr}_2$) in the def2-SVP basis set~\cite{weigend2005balanced}. For tractability, we truncate the $\text{Cr}_2$ Hilbert space via adaptive sampling configuration interaction (ASCI)~\cite{tubman2016deterministic}, selecting 4000 Slater determinants with the largest coefficients. Due to strong electron correlation, $\text{Cr}_2$ is a stringent test for evaluating GSE methods~\cite{larsson2022chromium,roos2003ground}.   In \cref{app:supplementary_figs}, we show results for lithium hydride (LiH) in the 3-21g basis set~\cite{binkley1980self}.

To simulate molecular data via (\ref{eqn:doverlapnoisy}), a key ingredient is an $N \times N$  Hamiltonian matrix $H$.  We begin with the ASCI Hamiltonian matrix $\overline{H}$.  In order to satisfy an ODMD convergence criterion~\cite{shen2023estimating}, we apply a simple linear transformation to compute $H = \beta_0 I + \beta_1 \overline{H}$.  We detail and justify this transformation in \cref{app:scaling}.

In this work, we use the time step $\Delta t = 1.0$.  Let $| \psi_n \rangle$ denote the $n$-th eigenvector of $H$; note that $H$ and $\overline{H}$ share the same eigenvectors.  For $p_0 \in (0,1)$, we prepare a reference state with ground state overlap $p_0$ via
\begin{align}
    |\phi_0\rangle = \sqrt{p_0} |\psi_0 \rangle + \sum_{n=1}^{N-1} \sqrt{\frac{1-p_0}{N-1}} |\psi_n \rangle.
\end{align}
With $H$, $\Delta t$, and $|\phi_0\rangle$ in hand, we compute the signal (\ref{eqn:doverlap}) from $k=0$ to $k=\Kmax$.  Taking the real part of the signal and adding noise with \emph{standard deviation} $\epsilon$, we obtain (\ref{eqn:noisyoverlap1}).  When constructing the Hankel (ODMD) and block Hankel (FDODMD) data matrices from \cref{sec:methods}, we set the delay parameter to $\rmd = \floor*{(K + 1)/2}$.    Using any of these methods, once we produce an estimate $E_0^\text{est}$ of the GSE of $H$, we apply the inverse of the above linear transformation to produce $( E_0^\text{est} - \beta_0) / \beta_1$, an estimate of the GSE of the true molecular system.


\textbf{Testing Protocol.} Once we have produced a full data trajectory $\mathcal{D} = \{ d(t_k) \}_{k=0}^{\Kmax}$, we choose $K$ such that $K+\rmd \leq \Kmax$ and restrict attention to $\mathcal{D}_K = \{ d(t_k) \}_{k=0}^{K+\rmd}$, the part of $\mathcal{D}$ needed to form the Hankel matrices $X, X'$ in (\ref{eqn:newxmat}-\ref{eqn:newxprimemat}) or their FDODMD counterparts. In this way, we simulate the situation in which we collect only $K+\rmd+1$ data points to begin with.  \emph{For each algorithm, we use only the data $\mathcal{D}_K$ to estimate the GSE and quantify the error between estimated and true GSEs.   Repeating this for the increasing sequence $K \in \{5, 10, 15, \ldots\}$, we quantify convergence of the GSE as a function of $K$, for each algorithm.}  For simplicity, in the rest of this paper, we refer to $K$ as the \emph{data length.} 

\section{Motivating Example}
\label{sec:depolarizing}
To further motivate the use of ODMD, we examine the efficacy of both ODMD and classic DFT-based spectral estimation on data corrupted with two forms of error.  In addition to shot noise, we account for global gate error accumulated during time evolution. The most applicable and analytically tractable model for such error is the depolarizing channel, which captures the stochastic state relaxation to the maximally mixed state with certain probability, hence erasing quantum information~\cite{dur2005standard}. This erasure noise in turn leads to an exponential damping of the sinusoidal time signal,
\begin{equation}\label{eqn:depolarizing}
    d_{\theta}(t) = e^{-\theta t}\langle \phi_0|e^{-iHt} |\phi_0\rangle + \xi(t), \quad \theta \in \mathbb{R},
\end{equation} 
where $\theta$ characterizes the strength of the global depolarizing channel. The exponential decay reflects the loss of quantum coherence over time and sets a practical limit on the timescale over which high-fidelity measurements can be collected from a quantum device. Furthermore, since the depolarizing noise channel effectively describes the cumulative impact of device-level errors as a uniform attenuation of the signal, it provides a useful benchmark for systematically evaluating and comparing quantum hardware performance under realistic conditions.

In general, the DFT constitutes a natural and convenient approach to extract energy information by decomposing a time domain signal into different frequency components, which reflect the eigenstate contributions encoded in the reference state. In noisy settings, however, spectral broadening can obscure the sharp frequency peaks, hence making it difficult to reliably estimate the GSE. This effect is particularly problematic when the evolution time is short, as spectral leakage from limited frequency resolution only exacerbates the challenge of isolating individual spectral features.

We illustrate this through an extended example.  To begin, we generate data $\mathcal{D} = \{ \Re d_{\theta}(k \Delta t) \}_{k=0}^{\Kmax}$ using the model (\ref{eqn:depolarizing}) with Gaussian shot noise $\xi(t) \sim  \mathcal{N}(0, \epsilon=0.01)$, depolarizing error with $\theta = 0.05$, reference state $|\phi_0\rangle$ with overlap $p_0 = 0.2$, time step $\Delta t = 1.0$, and $\Kmax = 300$.  All other details of data generation are as described in \cref{sec:data}.

We plot in \cref{fig:depolarized_spectrum} the magnitude of the DFT of the generated time series.  The DFT of a real sequence is symmetric about the zero frequency.  For positive frequencies, we see that there is no single dominant peak, as we would have obtained had we switched off both shot noise and depolarizing error. The spectrum shows broad envelopes in which signal and noise intermingle, complicating  GSE identification.

Continuing with the data $\mathcal{D}$, we study the convergence of the GSE estimation error incurred by two algorithms, baseline ODMD and a simple DFT-based method.  By baseline ODMD, we mean ODMD from \cref{subsec:ODMD} with no frequency domain denoising or signal stacking.  In the DFT-based method, we identify the discrete frequency bin $f_m$ that maximizes the absolute value of the DFT $| \hat{d}_{\theta}(f_m)|$. 

We apply the testing protocol described in \cref{sec:data} and plot in the left panel of \cref{fig:depolarizing_convergence} the absolute error of the estimated GSE as a function of data length $K$.  ODMD achieves rapid and steady convergence to chemical accuracy (dashed line) within short real-time evolution. DFT-based spectral estimation fails to converge; it cannot be used by itself for GSE estimation from noisy data.  As we will see in \cref{sec:results}, DFT-based spectral denoising \emph{in concert with ODMD} yields superior GSE estimates.


In addition to GSE estimation, ODMD also enables direct estimation of the depolarizing noise strength. Expanding (\ref{eqn:depolarizing}), we find
 \begin{equation}
     d_{\theta}(t) =\sum_{n=0}^{N-1}p_ne^{-(\theta + iE_n)t} + \xi(t).
 \end{equation}
When we apply ODMD to the time series $\{ d_{\theta}(t_k) \}$, the eigenvalues of the estimated propagator $A^\ast$ will be of the form $\tilde{\theta} + i\tilde{E}_n$.  As in (\ref{eqn:gseest}), the imaginary part contains our estimate of $E_n$, while the real part provides an estimate of $\theta$.  We apply this idea in conjunction with the testing protocol from \cref{sec:data}.  This yields the right panel of \cref{fig:depolarizing_convergence}: in the main plot, we plot the absolute error between estimated ($\tilde{\theta}$) and true ($\theta$) depolarizing parameters as a function of data length $K$ for the same experiment in the left panel.  Now suppose that we take the noisy data $\{ d_{\theta}(t_k) \}$, apply  the frequency domain denoising from \cref{sect:denoising}, and then multiply by $e^{+\tilde{\theta} t_k}$ to reverse the depolarization error and thus produce an estimate $\{ r(t_k) \}$ of the original underlying signal $\{ s(t_k) \}$.  In the inset of the right panel of \cref{fig:depolarizing_convergence}, we plot the original signal $s$, the noisy data $d$, and our estimate $r$.  At short times, the reconstructed signal $r$ well approximates the noiseless signal $s$. Deviations at longer times arise due to noise corruption, which we deal with in the next section.

\section{Results}
\label{sec:results}
Here we compare and evaluate FDODMD, ODMD, and DFT-based approaches for estimating the GSE of Cr$_2$ under shot noise.  Of main concern is the error between estimated ($\tilde{E}_0$) and true ($E_0$) molecular ground state energies \emph{as a function of data length $K$}.  \emph{To compute all convergence results below, we followed the testing protocol described in \cref{sec:data}.}  Our aim is to achieve chemical accuracy, defined as $|\tilde{E}_0 - E_0| < 10^{-3}$ Hartrees (units of energy). This tolerance reflects the  precision required for molecular predictions to be chemically meaningful.  In the plots below, this accuracy is indicated by black dashed lines.

We also investigate the sensitivity of all methods with respect to hyperparameters. These include $\delta$ (the SVD threshold chosen for FDODMD and ODMD) and $p_0$ (the overlap between the reference state and the true ground state).  We are also interested in investigating the convergence of all methods in the high noise regime $(\epsilon \geq 0.1)$.

\begin{figure}[t]
    \includegraphics[width=1.0\linewidth]{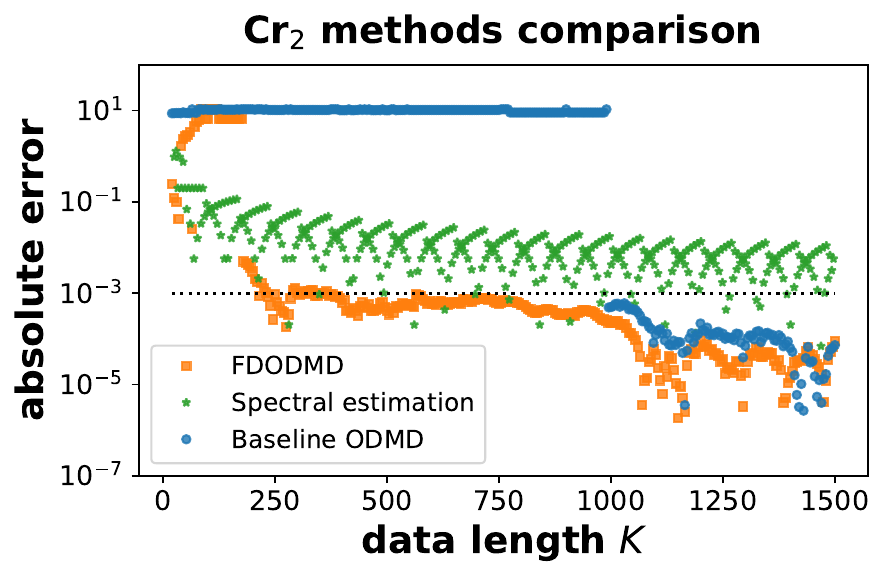}
    \caption{For a Cr$_2$ problem setting (low overlap $p_0=0.2$ and moderate noise standard deviation $\epsilon=0.1$) where both baseline ODMD (circles) and DFT-based spectral estimation (stars) fail to converge (e.g., for data length $K \leq 1000$), we see that FDODMD (squares) converges.  Convergence is defined as estimating the GSE to within $10^{-3}$ (dashed line) of its true value.  Data and algorithm parameters are detailed in \cref{sec:accelbaseline}.}
    \label{fig:fdodmd_vs_odmd_chem_acc}
\end{figure}

\begin{figure*}[ht]
    \centering
    \includegraphics[width=\linewidth]{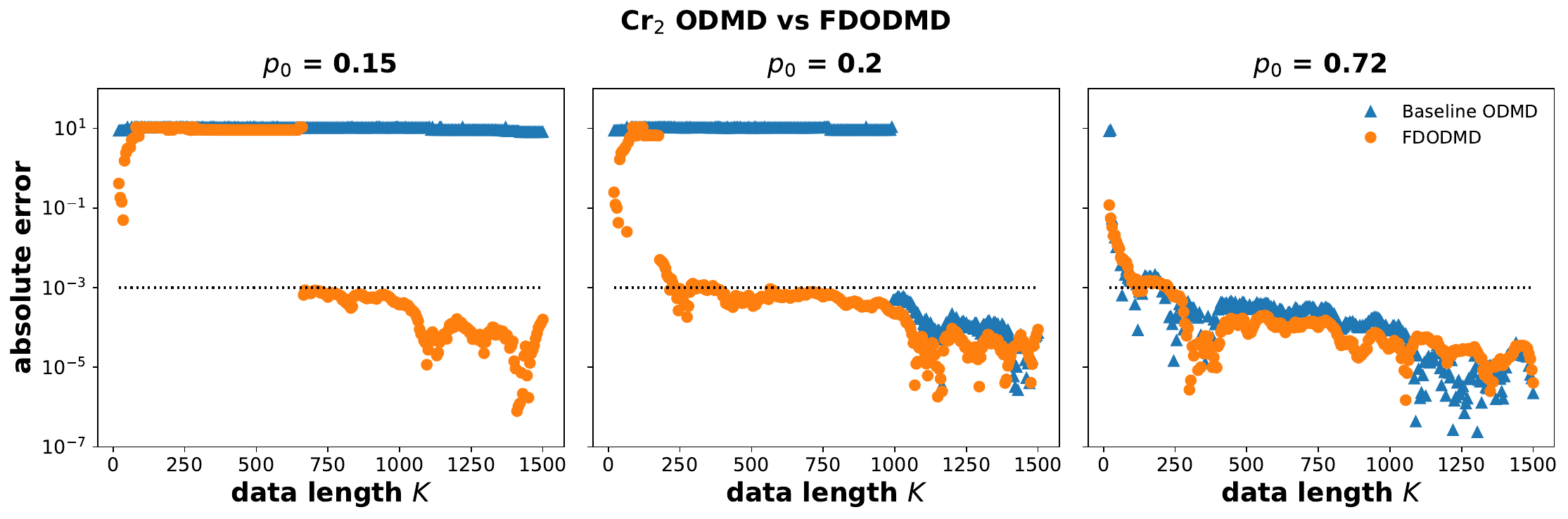}
    
    \caption{Convergence comparison of FDODMD (circles) and baseline ODMD (triangles) for Cr$_2$. We choose $\Kmax = 1500$, $\Delta t = 1.0$, and $p_0 \in\{0.15, 0.2, 0.72\}$. Both methods process noisy data given by (\ref{eqn:noisyoverlap1}) with $\xi(t_k) \sim \mathcal{N}(0, \epsilon=0.1)$. For FDODMD, we  use (\ref{eqn:dift}) with $\gamma \in \{1.0, 1.5, 2.0, 2.5, 3.0, 3.5\}$ to generate 6 denoised realizations. All the denoised realizations are then stacked in conjunction with the noisy data according to (\ref{eqn:signalstacked}). For both FDODMD and ODMD, we set $\delta = \epsilon$. Note the accelerated convergence to chemical accuracy offered by FDODMD for the lower overlap cases.}
    \label{fig:fdodmd_vs_odmd_chem_acc_w_overlap}
\end{figure*}

\begin{figure*}[ht]
    \centering
    \includegraphics[width=\textwidth]{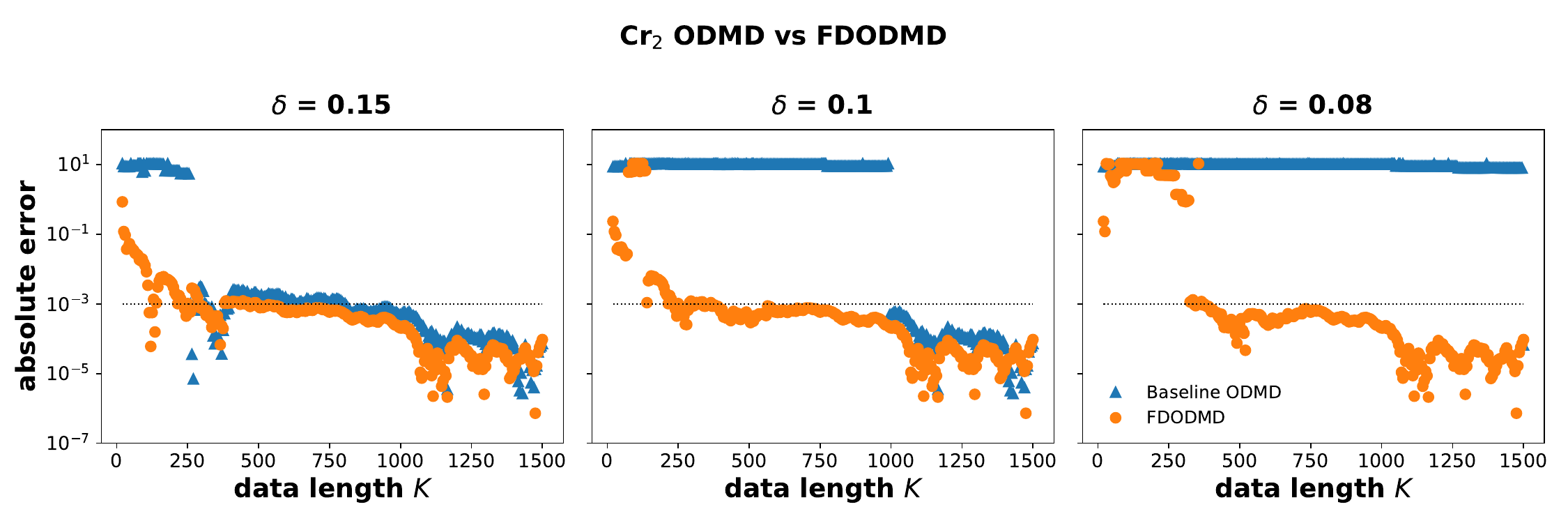}
    
    \caption{Robust invariance to SVD tolerance using FDODMD (squares) compared to SVD tolerance dependent ODMD (circles) for Cr$_2$. We choose $\Kmax = 1500$, $\Delta t = 1.0$, and $p_0 =0.2$. For both FDODMD and ODMD, we vary the SVD threshold $\delta \in \{0.15, 0.1, 0.08\}$. Both methods process noisy data given by (\ref{eqn:noisyoverlap1}) with $\xi(t_k) \sim \mathcal{N}(0, \epsilon=0.1)$. For FDODMD, we  use (\ref{eqn:dift}) with $\gamma \in \{1.0, 1.5, 2.0, 2.5, 3.0, 3.5\}$ to generate 6 denoised realizations. All the denoised realizations are then stacked in conjunction with the noisy data according to (\ref{eqn:signalstacked}).}
    \label{fig:fdodmd_vs_odmd_chem_acc_w_tol}
\end{figure*}

\subsection{Accelerated convergence relative to baseline methods}
\label{sec:accelbaseline}
We begin with a comparison of FDODMD, baseline ODMD, and DFT-based spectral estimation.  We simulate noisy data (\ref{eqn:noisyoverlap1}) for Cr$_2$ with $\Kmax = 1500$, $\Delta t = 1.0$, overlap $p_0=0.2$, and noise standard deviation $\epsilon = 0.1$.   We then apply the testing protocol from \cref{sec:data} to study the convergence (as a function of data length $K$) of three GSE estimation methods: (i) ODMD applied directly to the noisy data, (ii) simple DFT-based spectral estimation (described in \cref{sec:depolarizing}), and (iii) FDODMD.  For FDODMD, we  use the noisy data to generate $R=6$ denoised trajectories using (\ref{eqn:dift}) with thresholds $\gamma \in \{1.0, 1.5, 2.0, 2.5, 3.0, 3.5\}$.  We stack the noisy data with these denoised realizations and, using the procedure from \cref{sect:modmd}, compute the FDODMD estimate of the GSE.  For FDODMD and ODMD, we use the SVD threshold $\delta = \epsilon$.  

Although effective in many scenarios,  baseline ODMD alone cannot completely overcome difficulties posed by noise and, moreover, hyperparameter sensitivity. \Cref{fig:fdodmd_vs_odmd_chem_acc} illustrates these limitations. In certain regimes (i.e. noise standard deviation $\epsilon$, reference state overlap $p_0$, SVD threshold $\delta$), basic spectral estimation fails to converge (and is thus omitted from comparisons in 
\cref{fig:fdodmd_vs_odmd_chem_acc_w_overlap,fig:fdodmd_vs_odmd_chem_acc_w_tol,fig:combined_fd_convergence}), while ODMD requires large amounts of data (here $K \geq 1000$) to converge reliably. This reveals sensitivity to hyperparameter choice and increased quantum computational cost, both of which are undesirable. By contrast, our contribution in this paper, FDODMD, achieves chemical accuracy with faster convergence and improved robustness.

Since the algorithmic performance in the denoised framework is determined by the overall signal subspace rather than any individual time series, we can simply sweep over a range of thresholds $\{\gamma_i\}$ without needing to fine-tune a specific value. A distinctive advantage of our approach is that all denoised realizations $\{r_i(t_k)\}$ are generated classically from a single noisy observable trajectory, incurring no additional quantum cost, in circuits or measurements.

\subsection{Accelerated convergence with increased hyperparameter robustness}
Next we compare FDODMD and baseline ODMD for Cr$_2$, using reference states with varying levels of overlap $p_0$ with the true ground state; for both methods, we use the SVD threshold $\delta = \epsilon$.  We use the same data generation parameters as in \cref{sec:accelbaseline}.  Having already generated a trajectory with reference state overlap $p_0 = 0.2$, we generate two additional trajectories with respective overlaps $p_0 = 0.15$ and $p_0 = 0.72$; for each new trajectory, we denoise via (\ref{eqn:dift}) with thresholds given in \cref{sec:accelbaseline}.  This yields six denoised realizations for each of three noisy data time series.

We see from \cref{fig:fdodmd_vs_odmd_chem_acc_w_overlap} that in the low ground state overlap regime (e.g. $p_0 < 0.5$), FDODMD consistently outperforms ODMD in convergence behavior. ODMD either  requires a large data length (often exceeding $1000$) or even fails to reach chemical accuracy with maximal data length $\Kmax$. In contrast, FDODMD not only converges across all tested reference states but also does so significantly faster, requiring approximately $2.5$ times less quantum data than ODMD. This highlights the quantum resource savings enabled by denoising and stacking. For high ground state overlap ($p_0 = 0.72$), both approaches yield comparable convergence results. While quantum algorithms typically perform better with reference states that have a larger ground state overlap with the ground state, preparing such states can be costly~\cite{lin2020near}. The corresponding plot for LiH is shown in \cref{app:supplementary_figs}, \cref{fig:fdodmd_vs_odmd_chem_acc_w_overlap_LiH}.
\begin{figure}

  \centering
  \includegraphics[width=\linewidth]{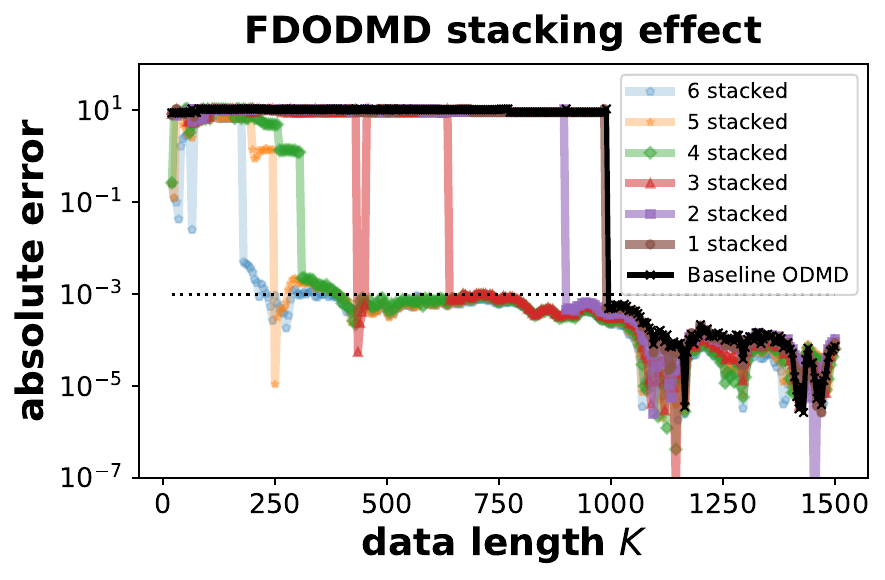}

\caption{Stacking with more denoised realizations improves FDODMD convergence.  We set $(\Kmax, \Delta t, p_0) = (1500, 1.0, 0.2)$ and apply Gaussian shot noise $\mathcal{N}(0, \epsilon=0.1)$ to simulate noisy data via (\ref{eqn:noisyoverlap1}). Denoised signals are then generated through (\ref{eqn:dift}) with $\gamma \in \{1.0, 1.5, 2.0, 2.5, 3.0, 3.5\}$. In the plot, \emph{1 stacked} corresponds to FDODMD where we stack the noisy data with $R=1$ denoised realization ($\gamma = 1.0$); \emph{2 stacked} corresponds to FDODMD where we stack the noisy data with $R=2$ denoised realizations ($\gamma \in \{1.0, 1.5\}$), etc. Baseline ODMD is applied directly to the noisy data.}
\label{fig:combined_fd_convergence}
\end{figure}

The use of an SVD threshold $\delta$ is essential for both ODMD and FDODMD: without it (i.e. $\delta = 0$), our experiments failed to  converge.  To test the $\delta$-dependence of both methods, we generate a new realization of the noisy Cr$_2$ trajectory with $p_0 = 0.2$ described in \cref{sec:accelbaseline}.  For each $\delta \in \{0.15, 0.1, 0.08 \}$, we test the convergence of both ODMD and FDODMD.   The results, shown in \cref{fig:fdodmd_vs_odmd_chem_acc_w_tol}, provide evidence that  moderate variations in $\delta$ do not strongly affect the performance of FDODMD. This is in contrast to baseline ODMD, which is highly sensitive to the choice of $\delta$.  The observed hyperparameter insensitivity of FDODMD is a very practical advantage, as the optimal $\delta$ is typically difficult to determine in advance without full noise characterization.



\Cref{fig:combined_fd_convergence} investigates how the performance of FDODMD changes with the number of denoised trajectories. We generate a new noisy data realization with parameters $(\Kmax, \Delta t, p_0, \epsilon) = (1500, 1.0, 0.2, 0.1)$ for Cr$_2$, and sequentially stack the denoised trajectories generated with $\gamma \in \{1.0, 1.5, 2.0, 2.5, 3.0, 3.5\}$, assessing how FDODMD convergence improves as we include more trajectories. Black crosses display the convergence of baseline ODMD. For both ODMD and FDODMD, we choose $\delta = \epsilon$.  We find that stacking more denoised signals markedly accelerates convergence to chemical accuracy, even though all methods eventually converge in the large data limit. This improvement presumably stems from the expanded signal subspace, which more effectively captures the underlying dynamics.

\textit{In total, the results in \cref{fig:fdodmd_vs_odmd_chem_acc_w_overlap,fig:fdodmd_vs_odmd_chem_acc_w_tol,fig:combined_fd_convergence} demonstrate FDODMD's robustness to algorithmic hyperparameters, which allow more rapid and stable  GSE estimation compared to ODMD.}

\begin{figure}
    \centering
    \includegraphics[width=1.0\linewidth]{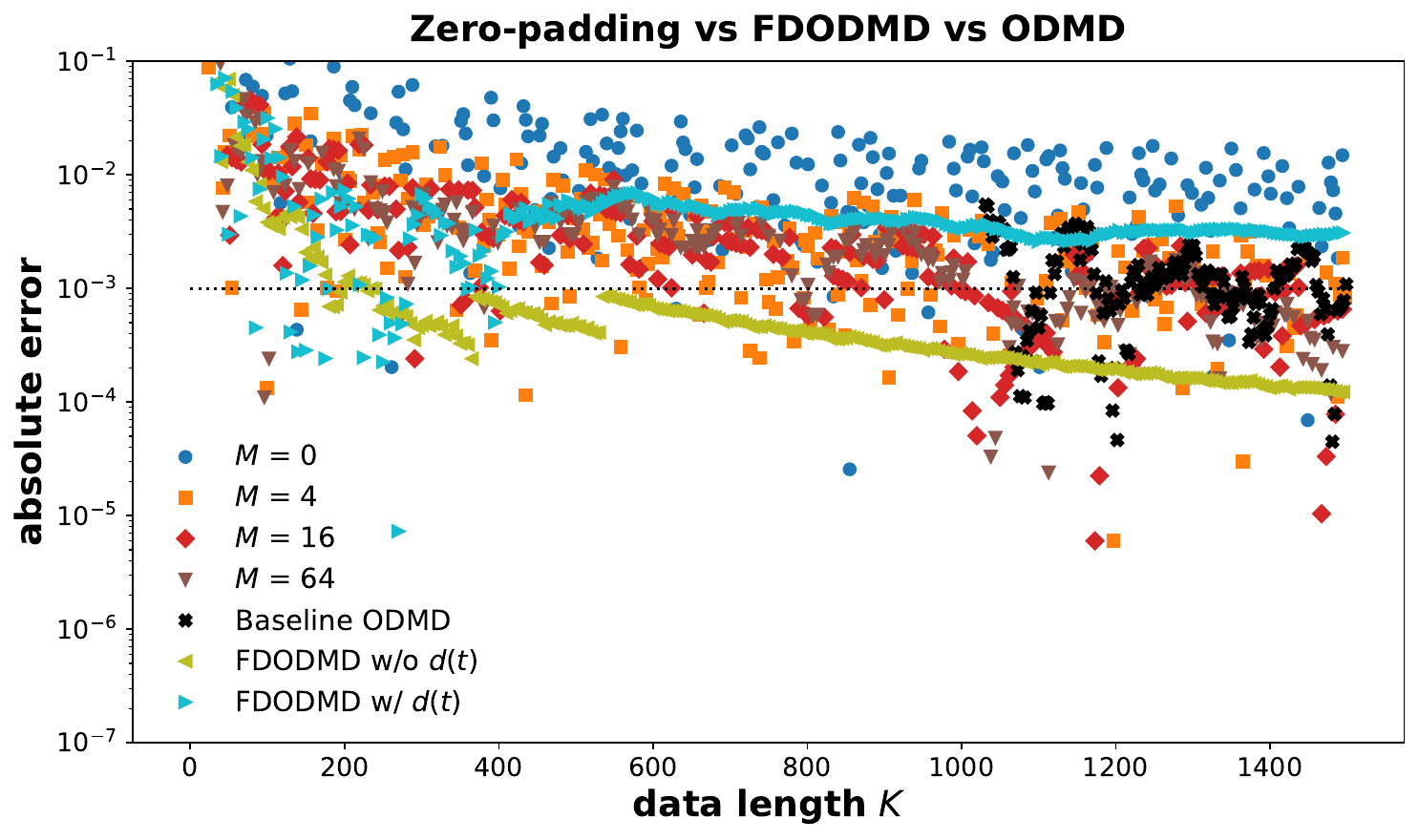}
    \caption{Convergence comparison of FDODMD, ODMD, and zero-padded DFT spectral estimation for the GSE of the Cr$_2$ molecule in the high noise regime ($\epsilon = 0.8$). For zero-padding, we append zeros equal to $M \in \{0,4,16,64\}$ times the length of the original time series, and perform DFT-based frequency estimation as in (\ref{eqn:zeropaddeddft}).  For both FDODMD and ODMD, we choose $\delta =\epsilon$ as the SVD parameter.  We generate noisy data using (\ref{eqn:noisyoverlap1}) with $\xi(t_k) \sim \mathcal{N}(0, \epsilon=0.8)$ and $(\Kmax, \Delta t, p_0) = (1500, 1.0, 0.2)$. 
    For FDODMD, we use (\ref{eqn:dift}) with $\gamma \in \{2.0, 2.5, 3.0, 3.5, 4.0, 4.5\}$ to generate $R=6$ denoised realizations.  FDODMD (excluding noisy data from the stack) achieves the most stable and accurate convergence; in this high noise regime, stacking \emph{with} the noisy data is detrimental for GSE estimation.}
    \label{fig:fdodmd_vs_odmd_vs_zeropad}
\end{figure}
\begin{figure}
    \centering
    \includegraphics[width=1.0\linewidth]{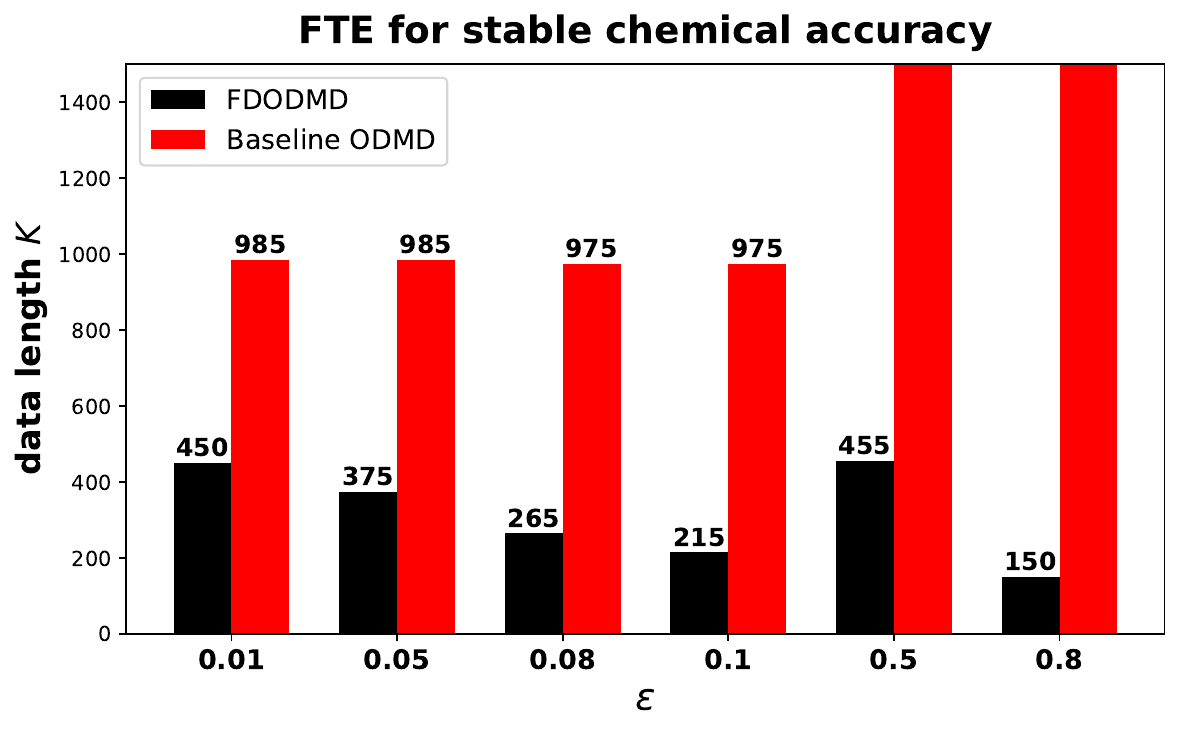}
    \caption{Stability of FDODMD under increasingly higher noise.  For each value of $\epsilon$, we generate noisy data using (\ref{eqn:noisyoverlap1}) with $\xi(t_k) \sim \mathcal{N}(0, \epsilon)$ and $(\Kmax, \Delta t, p_0) = (1500, 1.0, 0.2)$.  We report the data length required to reach stable chemical accuracy for Cr$_2$ molecule. ODMD processes the noisy data directly, whereas FDODMD uses a stack of $R=8$ denoised realizations constructed with $\gamma\in\{1.0, 1.5, 2.0, 2.5, 3.0, 3.5, 4.0, 4.5\}$.  The noisy data is excluded from the stack.  For all noise standard deviations $\epsilon$, FDODMD uses fewer quantum resources (shorter bar). For the highest noise cases ($\epsilon=0.5, 0.8$), baseline ODMD does not converge stably to chemical accuracy with data length within $\Kmax$.}
    \label{fig:compare_barplot}
\end{figure}

\subsection{Convergence in the high noise regime}
In \cref{fig:fdodmd_vs_odmd_vs_zeropad}, we compare FDODMD, baseline ODMD and the zero-padding procedure described in \cref{sec:methods}. For Cr$_2$, we generate noisy data with parameters $(\Kmax, \Delta t, p_0) = (1500, 1.0, 0.2)$ and noise standard deviation $\epsilon = 0.8$; we vary the zero-padding factor $M \in \{0, 4, 16, 64 \}$ in (\ref{eqn:zeropaddeddft}), where $M = 0$ corresponds to standard DFT frequency estimation. While higher padding factors ($M = 16, 64$) allow zero-padding to outperform ODMD in convergence speed, FDODMD still achieves superior stability and accuracy, as noise is taken into consideration. In addition, we assess the impact of incorporating the original noisy data into the FDODMD stacking procedure. As illustrated in \cref{fig:fdodmd_eigenvalue_convergence} in \cref{app:supplementary_figs}, the convergence of the estimated GSE is evaluated for both cases. When the noisy data is included, the results exhibit clear deviations from standard bounds~\cite{pollak2019tight, mccaskey2019quantum}, thereby indicating the need to exclude it for stable convergence. This procedure is practically feasible as the classical cost of running FDODMD multiple times is cheap. It is important to note that the question of whether to include the noisy data in the stack is unique to the high noise regime ($\epsilon\geq 0.5$).

As the noise standard deviation $\epsilon$ increases, the stability of baseline ODMD deteriorates, particularly for short time evolution. In \cref{fig:compare_barplot}, we compare the performance of ODMD and FDODMD on Cr$_2$ molecule with varying $\epsilon$. For FDODMD, we use $R=8$ denoised trajectories with $\gamma \in \{1.0, 1.5, 2.0, 2.5, 3.0, 3.5, 4.0, 4.5 \}$, this time excluding the original noisy data.  This exclusion is motivated by the results shown in \cref{fig:fdodmd_vs_odmd_vs_zeropad}. For both methods, we fix $(p_0, \Delta t, \Kmax) = (0.2, 1.0, 1500)$ and report in \cref{fig:compare_barplot} the data length required to achieve chemical accuracy for at least 10 consecutive values of $K$ (data length). Here taller bars indicate longer data lengths (i.e. higher quantum costs). When noise is high ($\epsilon = 0.5, 0.8$),  baseline ODMD fails to stably reach chemical accuracy, as manifested by the bars that exceed the budget. FDODMD, in contrast,  requires $\leq 455$ time steps to converge. When noise is small enough ($\epsilon < 0.1$) for ODMD to converge, FDODMD also does so using notably less data --- substantiating the efficiency gains from our denoising and stacking procedures.

\begin{figure}
    \centering
    \includegraphics[width=1.0\linewidth]{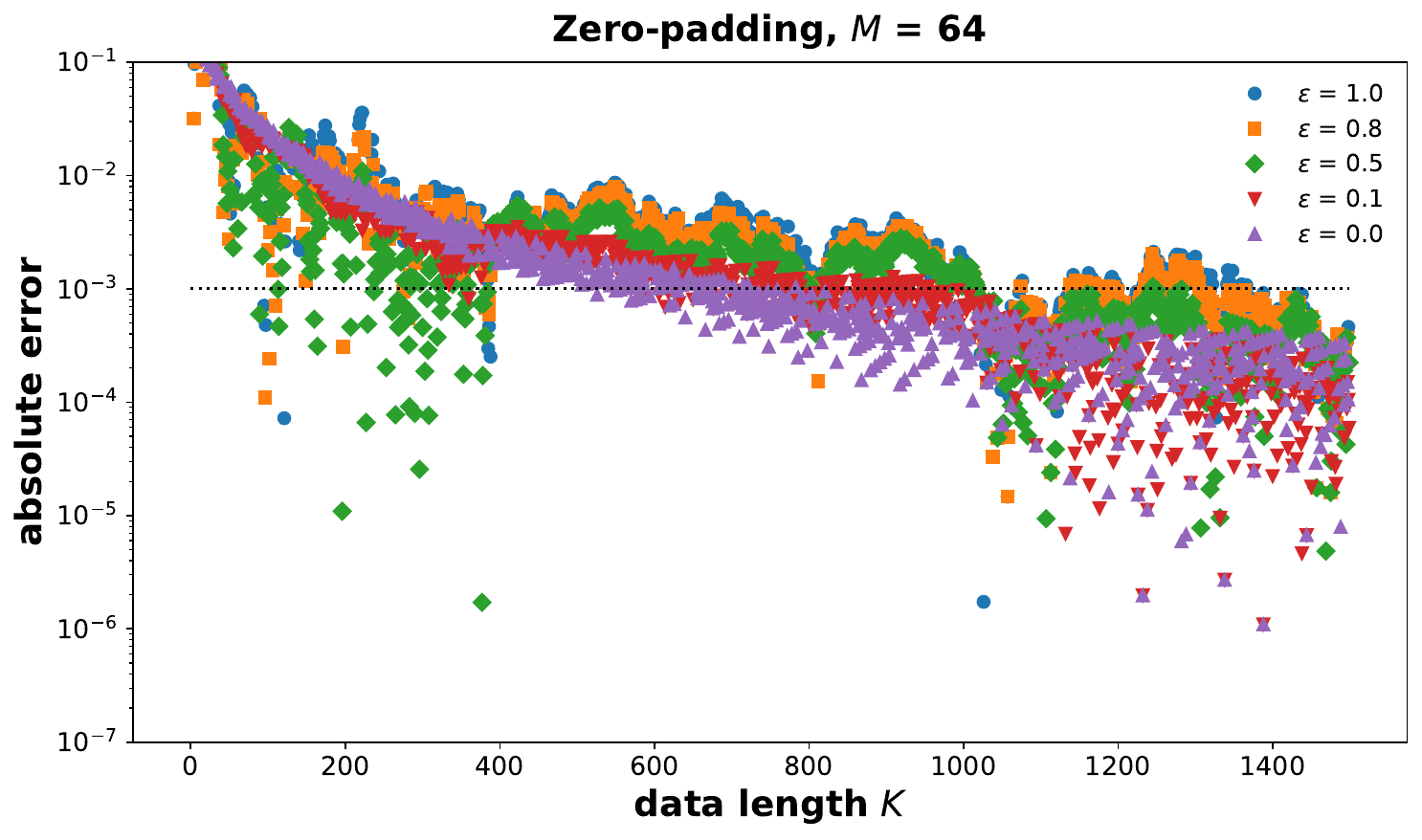}
    \caption{Noise robustness of zero-padding spectral estimation for the Cr$_2$ molecule. A fixed padding size $M=64$ is considered and the noise standard deviation $\epsilon$ are varied. For zero-padding, we append zeros equal to $M$ times the length of the original time series, and apply DFT-based frequency estimation as in (\ref{eqn:zeropaddeddft}).
   We generate noisy data using (\ref{eqn:noisyoverlap1}) with $\xi(t_k) \sim \mathcal{N}(0, \epsilon)$ and $(\Kmax, \Delta t, p_0) = (1500, 1.0, 0.2)$. 
    Overall, the convergence for a fixed padding factor is not strongly affected by variations in the noise standard deviation.}
    \label{fig:fixedpad}
\end{figure}

To further examine noise robustness, \cref{fig:fixedpad} studies the performance of the zero-padding procedure with a fixed padding factor $M=64$.  We apply this procedure to noisy trajectories with varying $\epsilon$.  In comparison to FDODMD and ODMD, the zero-padding method's overall error decay (as a function of data length) remains relatively insensitive to the noise standard deviation $\epsilon$.

\textit{In total, the results in \cref{fig:fdodmd_vs_odmd_vs_zeropad,fig:compare_barplot,fig:fixedpad} demonstrate that our methods deliver reliable convergence in the high noise regime, where baseline ODMD fails to converge within the available quantum resource budget.}

\section{Formal justification of FDODMD}\label{sec:formal_analysis}
\begin{figure*}[ht]
\subfloat[LiH denoising bound.\label{fig:lih_bound}]{%
\includegraphics[width=0.495\linewidth]{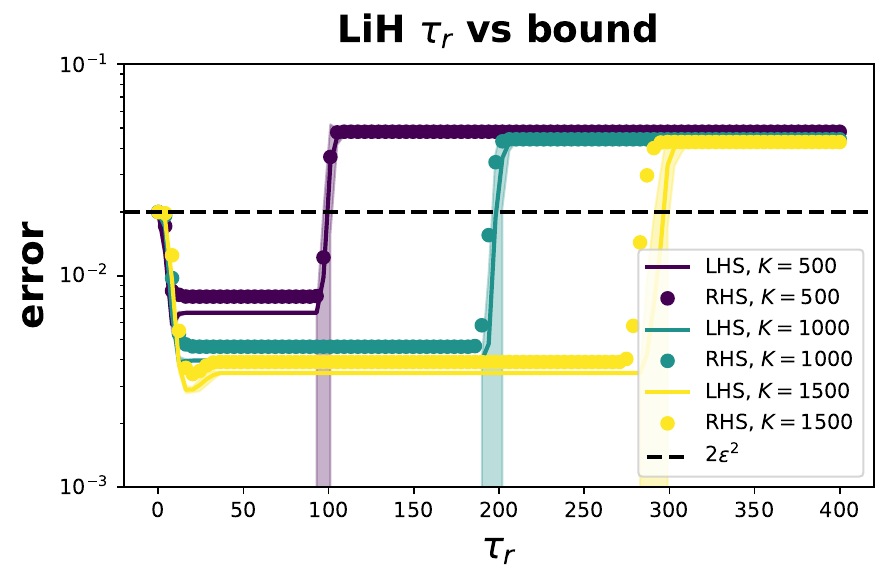}
}
\subfloat[Cr$_2$ denoising bound.\label{fig:cr2_bound}]{%
\includegraphics[width=0.495\linewidth]{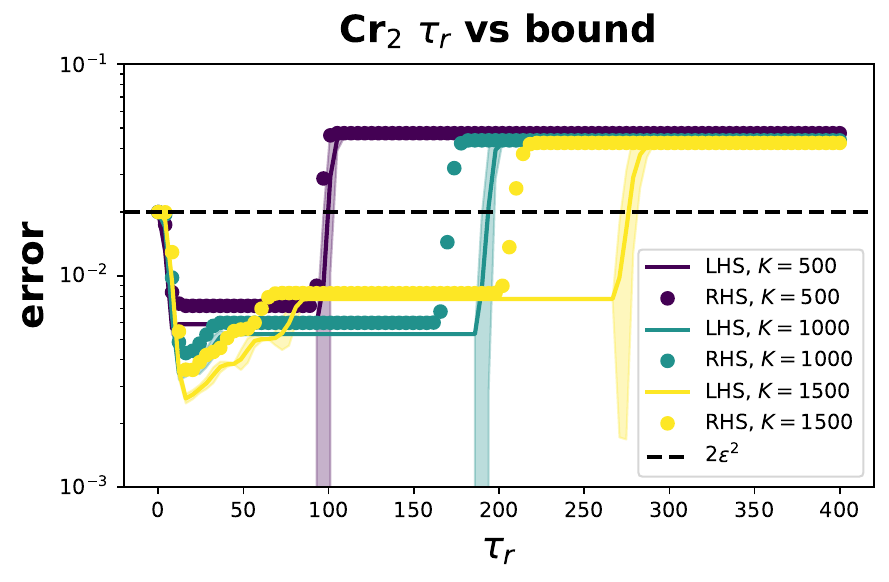}
}

\caption{Tightness of the bound developed in (\ref{eqn:combined}) for both LiH and Cr$_2$ averaged across 100 noisy trajectories. For the left-hand term (LHS), the mean is plotted in the solid line and the shaded band represents the standard deviation. This is not needed for the right-hand term (RHS) as it is deterministic. Here, for both molecular systems we fix $\epsilon=0.1$. Note that the $y$-axis here is rescaled by $1/K$. For three values of $K$, we see that the upper bound is reasonably close to the actual denoising error. As $\tau_r \to 0$, no thresholding/denoising is applied; thus both sides of the bound in (\ref{eqn:combined}) approach $2\epsilon^2$. As $\tau_r\to\infty$, the reconstructed signal vanishes ($\mathbf{r} \to 0$); thus both sides of the bound approach $\|\mathbf{s}\|_2^2$, the squared $2$-norm of the noiseless signal.}
\label{fig:lih_cr2_denoising_bound}
\end{figure*}
To support and complement the numerical results of \cref{sec:results}, we present an initial error analysis of FDODMD.

\subsection{Analysis of denoising}
In this subsection we offer error analysis of the denoising procedure detailed in \cref{sec:methods}. We defer all proof details to \cref{app:formal_proofs}. Let the noiseless, noisy, and denoised data be constructed by (\ref{eqn:doverlap}), (\ref{eqn:doverlapnoisy}), and (\ref{eqn:dift}), respectively; let $\mathbf{s}$, $\mathbf{d}$, and $\mathbf{r}$ denote their vectorized time domain representations.

\begin{theorem}\label{theorem:recon}
   The $L^2$ error between the reconstructed data $\mathbf{r}$ and noiseless data $\mathbf{s}$ is bounded above by
\begin{equation}\label{eqn:combined}
     \frac{1}{K} \E \left[ \| \mathbf{r} - \mathbf{s} \|_2^2 \right] \leq 2\epsilon^2 G(\tau_r,K, \epsilon, \mathbf{s} ),
\end{equation}
\end{theorem}
\noindent for a function $G$ that depends on the signal and noise characteristics. Here, we provide a proof outline; the full proof and an explicit functional form of $G$ can be found in \cref{app:formal_proofs}. Importantly, we validate that there exists, in general, threshold choices $\tau_r$ for which $(a)$ $G(\tau_r,K, \epsilon, \mathbf{s}) < G(0, K, \epsilon, \mathbf{s}) \equiv 1$ and $(b)$ $G$ diminishes as the length $K$ of the data increases.  This allows us to effectively denoise and accurately reconstruct the noiseless signal in the large $K$ limit.

\textit{Proof outline.} Direct calculation shows that the error between the noisy and noiseless data is
\begin{equation}
\label{eqn:dminuss}
\E \left[ \| \mathbf{d} - \mathbf{s} \|_2^2 \right] = \E \left[ \sum_{n=0}^{K-1} | \xi_n |^2 \right] = 2K \E[ (\Re \xi_0)^2 ] = 2 K \epsilon^2,
\end{equation}
where $\xi_n$ are noise random variables whose real and imaginary parts are i.i.d. (independently and identically distributed) Gaussians, i.e., $\Re \xi_n \sim \mathcal{N}(0,\epsilon)$ and $\Im \xi_n \sim \mathcal{N}(0,\epsilon)$. Examining the error between the denoised and noiseless data, we see that by Plancherel's theorem,
\begin{equation}
  \label{eqn:rminussprelim2}
    \| \mathbf{r} - \mathbf{s} \|_2^2 = \frac{1}{K} \| \widehat{\mathbf{r}} - \widehat{\mathbf{s}} \|_2^2 = \frac{1}{K} \sum_{k=0}^{K-1} \I_{ |\widehat{d}_k| \geq  \tau_r } | \widehat{\xi}_k |^2 + \I_{ |\widehat{d}_k| <  \tau_r } | \widehat{s}_k |^2,
\end{equation}
where $\I$ denotes an indicator variable. One can show that $|\hat{\xi_k}|^2$ are exponentially distributed with rate parameter $\lambda =\epsilon^2K^{-1}/2$.  Evaluating the two terms on the RHS of (\ref{eqn:rminussprelim2}), we obtain the bound $\E \left[ \| \mathbf{r} - \mathbf{s} \|_2^2 \right] \leq 2K\epsilon^2 G$ for a suitable function $G(\tau_r,K, \epsilon, \hat{s}_k)$, therefore completing the proof.

As $\tau_r \to 0$, we have $G \to 1$ so that both sides approach (\ref{eqn:dminuss}). As $\tau_r \to \infty$, both sides approach $\|\mathbf{s}\|_2^2$. Essentially, when no truncation is done, both sides of (\ref{eqn:combined}) approach the $2$-norm of the noise, and when a total truncation of all the modes is done, both side approach the $2$-norm of signal. 

Shown in \cref{fig:lih_cr2_denoising_bound} is a numerical comparison of the left and right-hand sides of (\ref{eqn:combined}) for both the molecular systems studied in this text. We fix the noise standard deviation $\epsilon=0.1$, time step $\Delta t = 1.0$, and initial state preparation with $p_0=0.2$, while varying the lengths $K$ of the observable series
and truncation threshold $\tau_r$. Numerically, we observe that the upper bound in (\ref{eqn:combined}) gives a reasonable estimate. In practice, since we set $\tau_r = \gamma_r *\text{median}(\hat{d}_n)$ with $\gamma_r \in [1,4.5]$, the $\tau_r$ values for the reported results lie within the range of $[2, 26]$. We find in \cref{fig:lih_cr2_denoising_bound} that this range is optimal. The large vertical jumps seen in the figure arise from cases where all frequency modes in the reconstruction are truncated.

\subsection{Analysis of the estimated GSE}
If $\overline{A} = X'X^+$ denotes the solution to (\ref{eqn:Asol}) using the noiseless versions of $X, X'$, then we can define $\tilde{A}=(X' + \delta X')(X+\delta X)^+$ to be the analogous solution using the denoised data, where $\delta X, \delta X'$ represent perturbations of $X, X'$, respectively. With this, it follows that
\begin{multline}
    \label{eqn:finalest}
\| \widetilde{A} - \overline{A} \|_2 \leq \frac{ \kappa }{ 1 - \kappa \eta} \left( \| \overline{A} \|_2 \eta + \| \rho \|_2 \frac{\kappa \eta}{\| X \|_2} +   \frac{\| \delta X' \|_2}{\| X \|_2} \right) \\+ \eta \| X \|_2 \| Y \|_2.
\end{multline}
Here $\kappa=\|X\|_2\| X^+\|_2$, $\eta=\|\delta X\|_2 / \|X\|_2$, $\rho = X'-AX$, and $Y=X'(X^\dagger X)^+$. The proof is a small perturbation of the arguments of Wedin \cite{wedin1973perturbation}, who considered a least squares problem of the form $a^\ast = \argmin_a \norm{x' - X a}_2$, where $a, x'$ are vectors and $X$ is a matrix.  Changing $a, x'$ to matrices $A, X'$, reversing the order of the product $X A$ to $A X$, and replacing Wedin's vector $2$-norm with the matrix Frobenius norm, one can repurpose Wedin's arguments\cite{wedin1973perturbation} to prove (\ref{eqn:finalest}).  From (\ref{eqn:finalest}), we see that the right-hand side tends to 0 as the perturbations $\delta X, \delta X'\to 0$. 

If $\tilde{A}_1$ represents the perturbed DMD solution using the denoised data, with corresponding perturbations $\delta X_1, \delta X_1'$ and $\tilde{A}_2$ represents the perturbed DMD solution using the noisy data, with corresponding perturbations $\delta X_2, \delta X_2'$, then knowing $\|\delta X_1\|_2 \leq \|\delta X_2\|_2$ and $\|\delta X_1'\|_2 \leq \|\delta X_2'\|_2$ is not sufficient to show that $\|\tilde{A}_1-\overline{A}\|_2 \leq \|\tilde{A}_2 - \overline{A}\|_2$. The bound in (\ref{eqn:finalest}) demonstrates that if $\|\tilde{A}_1-\overline{A}\|_2\leq C_1$ and $\|\tilde{A}_2-\overline{A}\|_2 \leq C_2$, then $C_1 \leq C_2$. Namely, this implies that denoising reduces an upper bound of the error between system matrices.

In practice, we know that the perturbations of the denoised data matrices and the noisy data matrices will obey $\|\delta X_1\|_2 \leq \|\delta X_2\|_2$ and $\|\delta X_1'\|_2 \leq \|\delta X_2'\|_2$. The bounds in (\ref{eqn:combined}) and (\ref{eqn:dminuss}) represent uniform upper bounds on the columns of $(\delta X_2, \delta X_2')$ and $(\delta X_1, \delta X_1')$, respectively. Thus, for correctly chosen truncation factors $\tau_r$, denoising reduces an upper bound on the error of the DMD system matrices. Since the eigenvalues of the DMD system matrices exactly determine the estimated GSE, we see that the denoising procedure reduces an upper bound on the error of our estimated GSE.

\section{Conclusion}\label{sec:conclusion}
In this work, we addressed the challenge of reliably estimating the ground state energy (GSE) of molecular systems from noisy quantum observables. Acknowledging that common noise sources such as depolarizing error and shot noise can significantly degrade the performance of many spectral estimation approaches, our goal was to develop a hybrid algorithm that remains robust to a high amount of noise while exhibiting a low sensitivity to its algorithmic hyperparameters.

We introduced a novel framework --- Fourier denoising ODMD (FDODMD) --- which builds on observable dynamic mode decomposition (ODMD) by combining classical Fourier-based thresholding with quantum signal subspace techniques. By leveraging frequency-domain thresholding to suppress spurious spectral modes introduced by noise, FDODMD reconstructs cleaner time-domain signals from a single noisy observable trajectory. It then generates multiple denoised realizations across varying thresholds and stacks them into a higher-dimensional signal subspace, enabling the extraction of richer spectral information. This approach incurs no additional quantum cost and significantly improves convergence to chemical accuracy in high error scenarios, even in regimes where baseline ODMD fails.



Numerical experiments on challenging molecular systems, such as LiH and Cr$_2$, demonstrate that FDODMD consistently outperforms baseline ODMD and classical Fourier spectral estimation. The results reflect rapid energy convergence with markedly shorter simulation times, substantially lowering the quantum resources required. Moreover, our analytical bounds validate that the denoising procedure effectively reduces the reconstruction error relative to the noisy dynamics.

In summary, FDODMD provides a robust and resource-efficient algorithm that overcomes some key obstacles of GSE estimations in noisy settings. By combining classical Fourier denoising with a quantum signal subspace eigensolver, it opens up a promising avenue for the design of accurate and scalable hybrid algorithms for computational chemistry and beyond.

\section*{Acknowledgments}
Research was sponsored by the Office of Naval Research and was accomplished under Grant Number W911NF-23-1-0153. The views and conclusions contained in this document are those of the authors and should not be interpreted as representing the official policies, either expressed or implied, of the Army Research Office or the U.S. Government. The U.S. Government is authorized to reproduce and distribute reprints for Government purposes notwithstanding any copyright notation herein. We acknowledge support from the U.S. Department of Energy, Office of Science, Accelerated Research in Quantum Computing, Fundamental Algorithmic Research toward Quantum Utility (FAR-Qu). H. Bassi acknowledges partial support from NSF DMS-1840265. This research used resources of the National Energy Research Scientific Computing Center (NERSC), a U.S. Department of Energy Office of Science User Facility located at Lawrence Berkeley National Laboratory, operated under Contract No.~DE-AC02-05CH11231, using NERSC award ASCR-m4577.
\section*{Data Availability}
The data that support the findings of this study are available at
\url{https://github.com/hbassi/denoisingodmd}.
\section*{Conflict of Interest}
The authors have no conflicts to disclose.   

\section*{Author Contributions}
H.B. - Methodology, software, validation, formal analysis, investigation, data curation, writing - original draft, writing - review \& editing, visualization

Y.S. - Conceptualization, methodology, validation, formal analysis, investigation,  writing - original draft, writing - review \& editing, visualization, supervision

H.S.B. - Conceptualization, methodology, formal analysis, investigation, writing - original draft, writing - review \& editing, supervision, project administration, funding acquisition

R.V.B. - Conceptualization, methodology,  writing - review \& editing, supervision, project administration, funding acquisition

\bibliographystyle{aipnum4-2}   
\bibliography{references.bib}

\appendix

\section{Optimal shot allocation}
\label{app:optimal_shot_allocation}

Suppose we ask: how should we allocate shots at each time $t_k$, from a fixed total budget of $N_{\text{total}}$ shots, so as to minimize the uncertainty in each noisy $\Re d(t_k)$ defined in (\ref{eqn:noisyoverlap2})?  A general answer to this question that assumes no prior knowledge of $\Re s(t_k)$ is beyond the scope of this paper. However, if we take a hindsight/oracle point of view, whereby $\Re s(t_k)$ is known and we want to determine after the fact how many shots we should have used to minimize uncertainty, then a simple solution is possible.

Let us quantify the total uncertainty by the total variance of $\Re d(t_k)$ over all times.  Summing (\ref{eqn:varshots}) over all $k$ and using $\pi_{t_k} = (1 + \Re s(t_k))/2$, we get $\sum_k \text{Var}(\Re d(t_k)) = \sum_k  (1 - \Re s(t_k)^2)/{n_{t_k}}$. This leads us naturally to the constrained optimization: \begin{equation}
\label{eqn:minvaropt1}
\begin{split}
\text{minimize} \quad & \sum_k (1 - \Re s(t_k)^2)/{n_{t_k}} \\
\text{subject to} \quad & \sum_k n_{t_k} = N_{\text{total}} \text{ and } n_{t_k} \geq 0 .
\end{split}
\end{equation} 
Here we minimize with respect to $n_{t_k}$, the number of shots to allocate at time $t_k$.  To solve this optimization problem, we use Lagrange multipliers, i.e., we seek critical points $(n_{t_k}^*, \mu^*)$ of the Lagrangian
\begin{equation}\label{eqn:lagvaropt}
    \mathcal{L}(n_{t_k}, \mu) = \sum_k (1 - \Re s(t_k) ^2)/{n_{t_k}} + \mu \Bigl( \sum_k n_{t_k} - N_{\text{total}} \Bigr).
\end{equation}
Setting $\nabla_{n_{t_k}}\mathcal{L}(n_{t_k}, \mu) = 0$ yields
$n_{t_k} = \sqrt{(1 - s(t_k)^2)/{\mu}}$.
Combining this with the constraint $\sum_k n_{t_k} = N_{\text{total}}$ enables us to solve for the optimal $\mu$. Substituting such optimal $\mu$ back into the $n_{t_k}$ equation yields the optimal number of shots to allocate at time $t_k$: 
\begin{equation}\label{eqn:nstarfinal}
    \floor*{n_{t_k}^\ast} = N_{\text{total}} \left( \frac{\sqrt{1 - s(t_k)^2}}{\sum_k \sqrt{1 - s(t_k)^2}} \right).
\end{equation}
Note that  (\ref{eqn:nstarfinal}) is minimized when $s(t_k) = \pm 1$, and maximized when $s(t_k) = 0$. 

\begin{figure}
    \centering
    \includegraphics[width=1.0\linewidth]{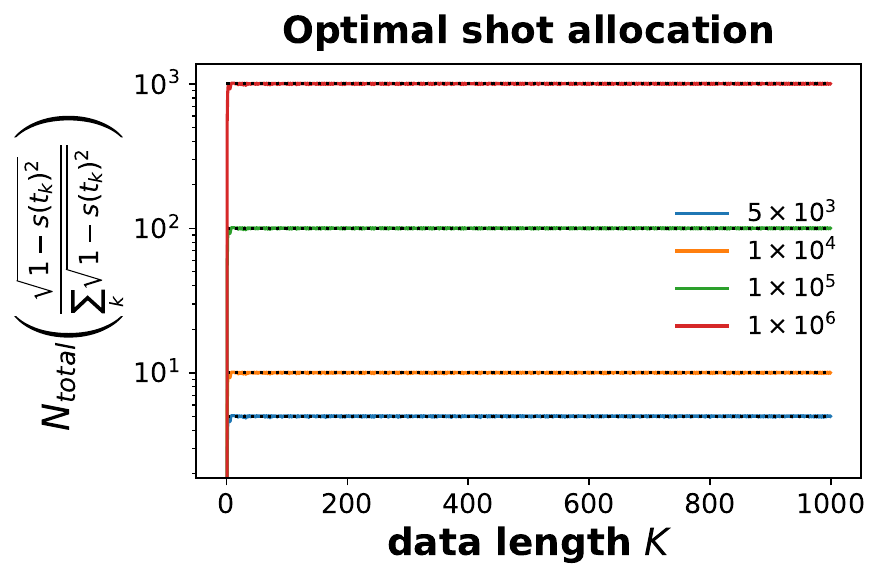}
    \caption{Optimal shot allocation for $N_{\text{total}} = \{5 \times 10^3, 10^4, 10^5, 10^6\}$. We compute (\ref{eqn:doverlap}) with $(\Kmax, \Delta t, p_0) = (1000, 1.0, 0.2)$ as input data to (\ref{eqn:nstarfinal}). We find that it is nearly optimal to allocate an equal number of shots across all time.}
    \label{fig:optimal_shot_allocation}
\end{figure}

We view (\ref{eqn:nstarfinal}) as justifying the use of an equal number of shots at each time $t_k$. In \cref{fig:optimal_shot_allocation}, we plot the optimal number of shots $n^\ast_{t_k}$ at time $t_k$, prescribed by  (\ref{eqn:nstarfinal}), for various total shot budgets $N_{\text{total}} \in \{5 \times 10^3, 10^4, 10^5, 10^6\}$. Dashed black lines indicate the uniform allocation $N_{\text{total}} / \Kmax$ with $\Kmax = 1000$. At time $t = 0$, $s(0) = \langle \phi_0 | \phi_0\rangle = 1$, implying zero variance --- thus no shots are needed at $t=0$.

For every $t_k > 0$, we find in \cref{fig:optimal_shot_allocation} that $n^\ast_{t_k}$, calculated with the noiseless observable $s(t_k)$ in (\ref{eqn:nstarfinal}), is either $N_{\text{total}}/\Kmax$ or $N_{\text{total}}/\Kmax - 1$. Empirically, using an equal number of shots $N_{\text{total}}/\Kmax$ at each time $t_k$ is essentially equivalent to optimally allocating the shots by minimizing the total variance; the gap in the performance between equal and optimal shot allocation is negligible.

\section{Hamiltonian Scaling}
\label{app:scaling}
Let $\overline{H}$ denote the molecular Hamiltonian whose GSE must be estimated; in this paper, this is the ASCI Hamiltonian.  Let $\{ \overline{E}_0, \ldots, \overline{E}_{N-1} \}$ denote the eigenvalues of $\overline{H}$; similarly, let $\{ E_0, \ldots, E_{N-1} \}$ denote the eigenvalues of a linearly rescaled Hamiltonian
\begin{equation}
\label{eqn:linrescale}
H = \beta_0 I + \beta_1 \overline{H}.
\end{equation}
We assume all eigenvalues are indexed in nondecreasing order.  In order to satisfy an ODMD convergence criterion~\cite{shen2023estimating}, we seek a rescaling that implies
\begin{equation}
\label{eqn:odmdbound}
(E_{N-1} + E_1 - 2 E_0) \Delta t < 2 \pi.
\end{equation}
Let $\underline{E}$ and $\overline{E}$ be \emph{estimated} lower and bounds, respectively, on the true spectrum; by this we mean that for all $j = 0, 1, \ldots, N-1$, we must have
\begin{equation}
\label{eqn:barcond}
\underline{E} < \overline{E}_j < \overline{E}.
\end{equation}
In a practical setting, $\underline{E}$ and $\overline{E}$ can be estimated as follows.  First one runs a Hartree-Fock calculation on the molecular system under consideration to calculate Hartree-Fock eigenvalues $\tilde{E}_0$ and $\tilde{E}_{N-1}$ that approximate the true eigenvalues with bounds~\cite{mcweeny1989method,pollak2019tight} of the form $| \tilde{E}_0 - \overline{E}_0 | < \alpha_0$ and $| \tilde{E}_{N-1} - \overline{E}_{N-1} | < \alpha_{N-1}$.  Let $\alpha = \max\{ \alpha_0, \alpha_{N-1} \}$ and set
\[
\underline{E} = \tilde{E}_0 - \alpha \ \text{ and } \
\overline{E} = \tilde{E}_{N-1} + \alpha.
\]
By construction, (\ref{eqn:barcond}) will be satisfied.  In the present paper, \emph{to simulate the above scenario}, we use the true ASCI eigenvalues with a large value of $\alpha$ (in particular, $\alpha = 0.2$) and set $\underline{E} = \overline{E}_0 - \alpha$ and $\overline{E} = \overline{E}_{N-1} + \alpha$.  Now define $\Delta E = \overline{E} - \underline{E}$ and $\mu_E = (\overline{E} + \underline{E})/2$.
We then use
\[
\beta_0 = -\pi \mu_E (2 \Delta t \Delta E)^{-1}  \ \text{ and } \ \beta_1 = \pi (2 \Delta t \Delta E)^{-1}. 
\]
Let us show that this rescaling satisfies (\ref{eqn:odmdbound}).  First, since $\overline{H}$ is Hermitian, there exists a unitary $V$ such that $\overline{H} = V \overline{\Lambda} V^\dagger$ where $\overline{\Lambda}$ is diagonal and contains the eigenvalues of $\overline{H}$.  Note that $H$ is also diagonalized by $V$; by (\ref{eqn:linrescale}), $V^\dagger H V = \beta_0 I + \beta_1 \overline{\Lambda}$,
which is purely diagonal.  Thus the eigenvalues of $H$ and $\overline{H}$ are related via
\[
E_j = \beta_0 + \beta_1 \overline{E}_j.
\]
Using this and (\ref{eqn:barcond}), we have
\begin{align*}
E_{N-1} - E_0 &= \frac{\pi}{2 \Delta t} \frac{ \overline{E}_{N-1} - \overline{E}_0 }{ \overline{E} - \underline{E}} < \frac{\pi}{2 \Delta t} \\
E_{1} - E_0 &= \frac{\pi}{2 \Delta t} \frac{ \overline{E}_{1} - \overline{E}_0 }{ \overline{E} - \underline{E}} < \frac{\pi}{2 \Delta t}
\end{align*}
Adding these inequalities and multiplying through by $\Delta t$, we see that (\ref{eqn:odmdbound}) is satisfied.  In the above derivation, $\beta_0$ canceled out entirely; the purpose of $\beta_0$ is to ensure that the rescaled eigenvalues satisfy $E_j \in [-\pi/(4 \Delta t), \pi/(4 \Delta t)]$.  To see this, note that the above definitions and (\ref{eqn:barcond}) yield
\begin{align*}
E_0 &= \frac{\pi}{2 \Delta t \Delta E}[ -\underline{E}/2 - \overline{E}/2 + \overline{E}_0] \\
 &> \frac{\pi}{2 \Delta t \Delta E}[ -\underline{E}/2 - \overline{E}/2 + \underline{E}] \\
 &= \frac{\pi}{2 \Delta t \Delta E} \left[-\frac{\Delta E}{2}\right] = -\frac{\pi}{4 \Delta t}
\end{align*}
Analogously, we can show $E_{N-1} < \pi / (4 \Delta t)$.  This together with the nondecreasing nature of the eigenvalues implies that all rescaled eigenvalues are in the desired interval.  When $\Delta t = 1.0$, as it is in this paper, we have $E_j \in [-\pi/4, \pi/4]$ for all $j$.

\section{Supplementary Figures}
\label{app:supplementary_figs}

 \begin{figure*}[ht]
    \centering
    \includegraphics[width=\linewidth]{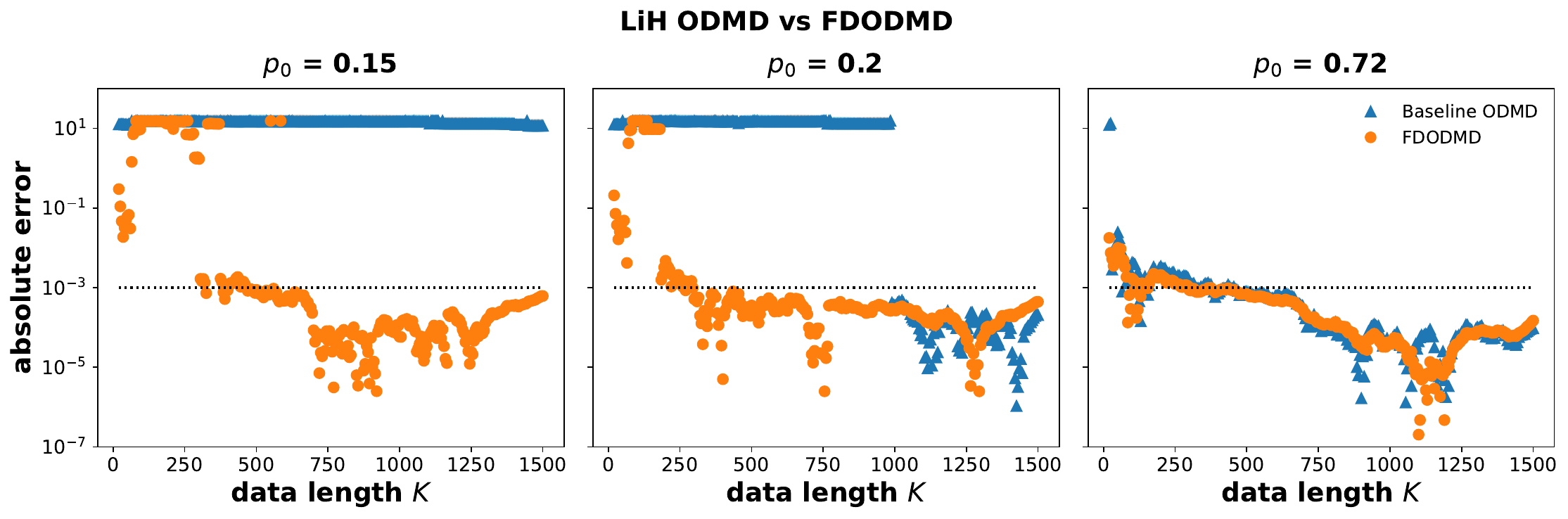}

    \caption{Convergence comparison of FDODMD (circles) and baseline ODMD (triangles) for LiH. We choose $\Kmax = 1500$, $\Delta t = 1.0$, and $p_0 \in\{0.15, 0.2, 0.72\}$. Both methods process noisy data given by (\ref{eqn:noisyoverlap1}) with $\xi(t_k) \sim \mathcal{N}(0, \epsilon=0.1)$. For FDODMD, we  use (\ref{eqn:dift}) with $\gamma \in \{1.0, 1.5, 2.0, 2.5, 3.0, 3.5\}$ to generate 6 denoised realizations. All the denoised realizations are then stacked in conjunction with the noisy data according to (\ref{eqn:signalstacked}). For both FDODMD and ODMD, we set $\delta = \epsilon$. Note the accelerated convergence (roughly $4$ times less data required) to chemical accuracy offered by FDODMD for the lower overlap cases.}
    \label{fig:fdodmd_vs_odmd_chem_acc_w_overlap_LiH}
\end{figure*}

\begin{figure}[htbp!]
    \centering
\includegraphics[width=1.0\linewidth]{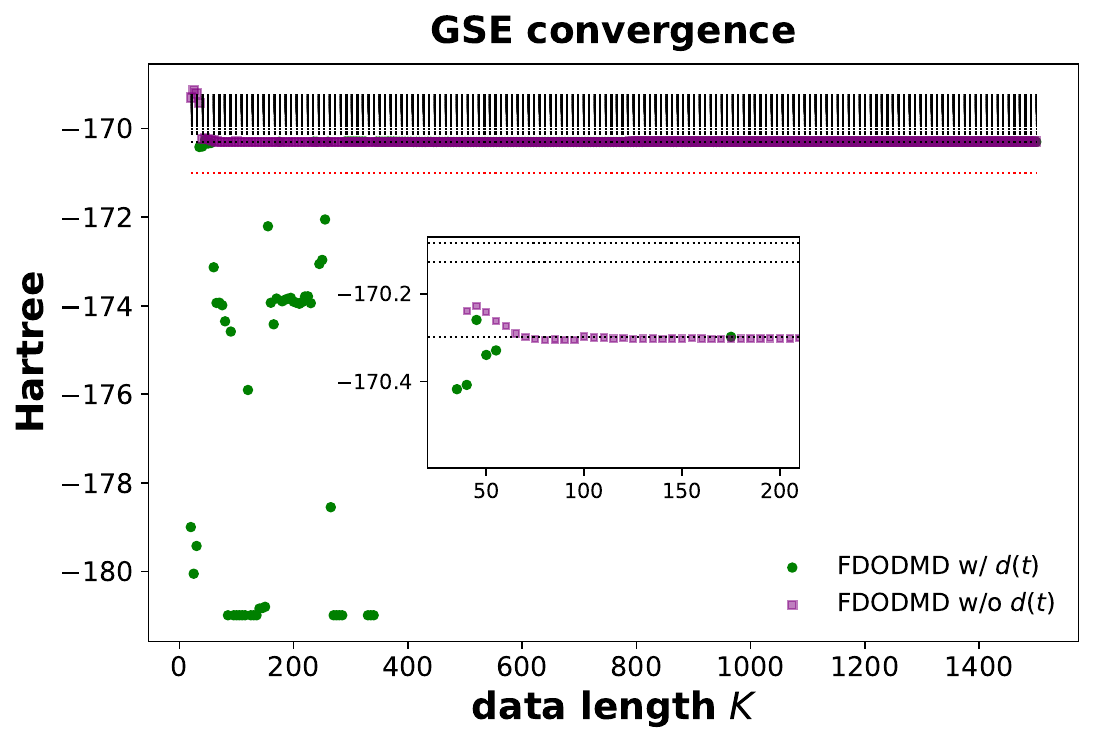}
\caption{Eigenvalue convergence of FDODMD with and without stacking the raw noisy data for Cr$_2$ molecule.  Dashed lines indicate the target eigenenergies of $H$. We fix $(\Kmax, \Delta t, p_0) = (1500, 1.0, 0.2)$ and simulate noisy data given by (\ref{eqn:noisyoverlap1}) with $\xi(t_k) \sim \mathcal{N}(0, \epsilon=0.8)$. We choose $\delta = 0.8$ accordingly. For FDODMD, we use (\ref{eqn:dift}) with $\gamma \in \{2.0, 2.5, 3.0, 3.5, 4.0, 4.5\}$ to generate $R=6$ denoised realizations. The red dashed line indicates the lower bound computed from Ref. \onlinecite{pollak2019tight}. When stacking includes the raw data in the high noise regime, the estimated ground state energy exhibits a clear violation of known lower bounds. }
    \label{fig:fdodmd_eigenvalue_convergence}
\end{figure}

\begin{figure*}[ht!]
\subfloat[LiH  \label{fig:lih_compare}]{%
\includegraphics[width=0.475\linewidth]{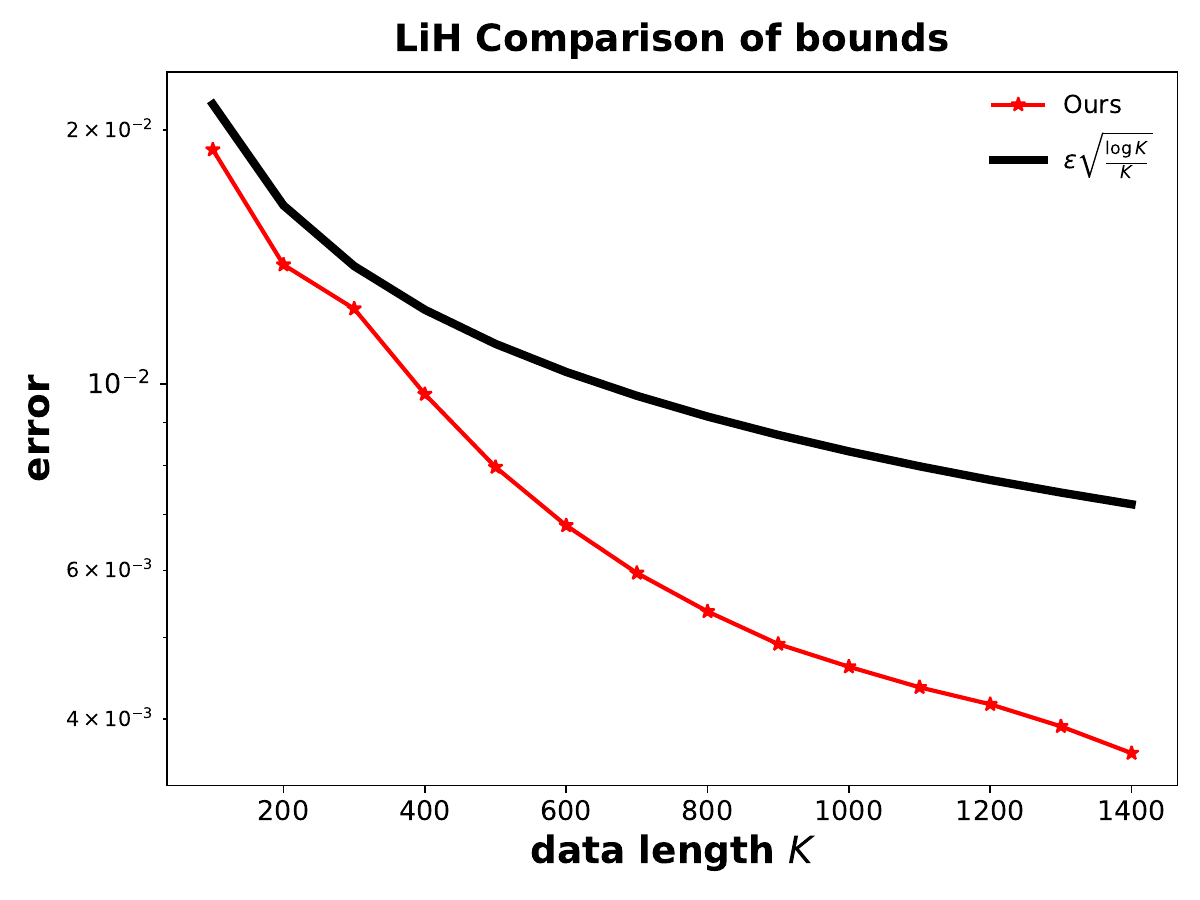}
}
\subfloat[Cr$_2$\label{fig:cr2_compare}]{%
\includegraphics[width=0.5\linewidth]{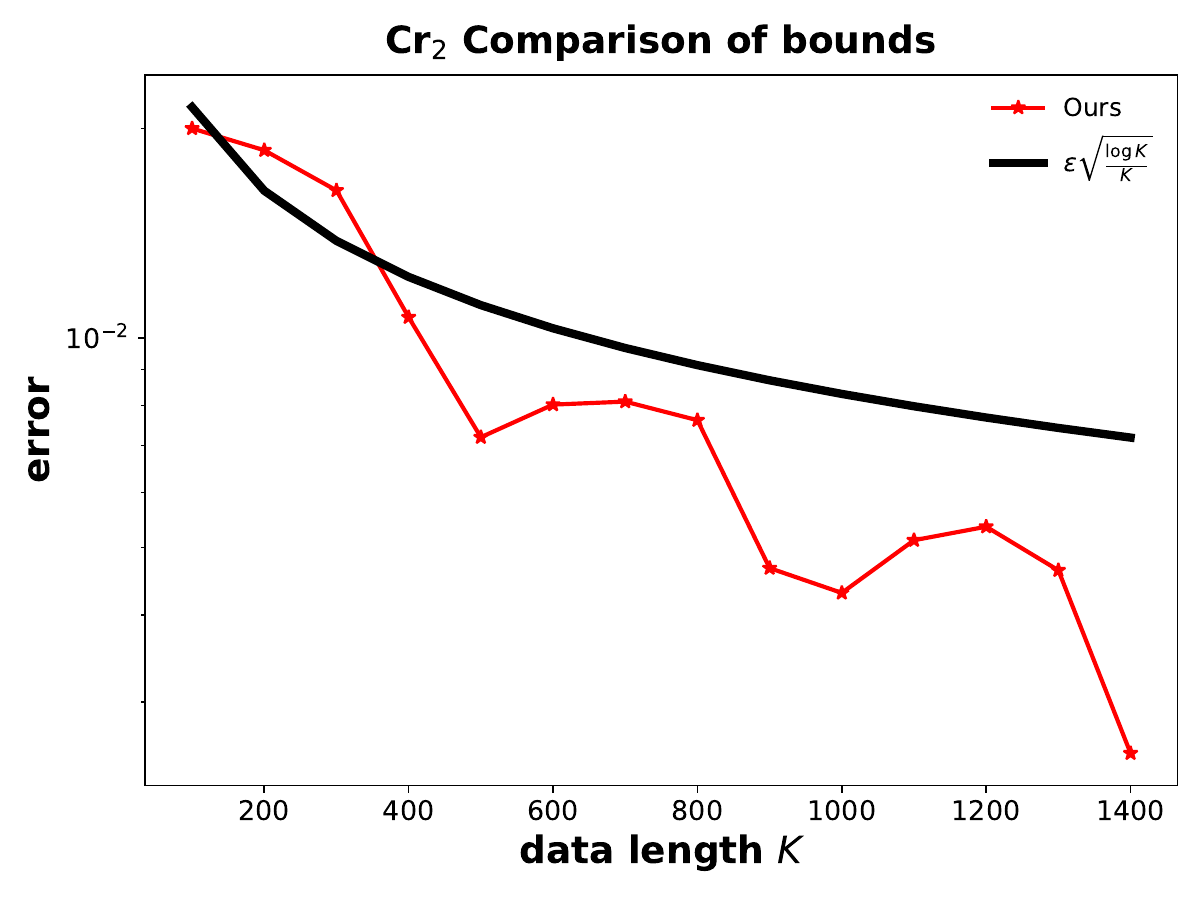}
}
\caption{Comparison of the bound developed in (\ref{eqn:combined}) for both LiH and Cr$_2$ averaged across 100 noisy trajectories compared to optimal bound in \cite{bhaskar2013atomic}. For our bound, we choose the optimal truncation factor $\tau_r$ consistent with \cref{fig:lih_cr2_denoising_bound}. Here, for both molecular systems we fix $\epsilon=0.1$, $p_0=0.2$, and $\Delta t = 1.0$ to simulate noisy  data from (\ref{eqn:noisyoverlap1}). We vary the length of the time series $K$ for both bounds. Numerically, our bound appears tighter for these experiments. }
\label{fig:lih_cr2_denoising_bound_compare}
\end{figure*}
\Cref{fig:fdodmd_vs_odmd_chem_acc_w_overlap_LiH} compares convergence to the GSE of another test molecule, LiH in the 3-21g basis, under the same parameter setting as in \cref{fig:fdodmd_vs_odmd_chem_acc_w_overlap}. We investigate the effect of varying the initial state overlap $p_0$. Similar to Cr$_2$, we find that FDODMD offers rapid and stable convergence compared to ODMD, requiring roughly $4$ times less quantum resources.


\Cref{fig:fdodmd_eigenvalue_convergence} examines the convergence of the FDODMD GSE estimate under the same parameter set used in \cref{fig:fdodmd_vs_odmd_vs_zeropad}.
We compare the effect of stacking with and without the raw noisy data $d(t)$ from (\ref{eqn:doverlapnoisy}) in the high noise regime. Specifically, we evaluate two stacking strategies:
\begin{enumerate}
\item Stacking the raw noisy observable $d(t)$ together with its $R$ denoised reconstructions.  
\item Stacking only the $R$ denoised trajectories.  
\end{enumerate}
As the denoising and stacking procedures are purely classical, both variants can be evaluated with minor cost by running the post-processing twice. When the noisy data is retained in the stack, the estimated GSE, as a function of the data length, manifestly violates known lower bounds reported in the literature~\cite{pollak2019tight,mcweeny1989method}. By contrast, the stack composed solely of the denoised signals respects both the lower and upper bounds over the observed time window.

\Cref{fig:lih_cr2_denoising_bound_compare} compares the error bound derived in (\ref{eqn:combined}) with an alternative denoising bound based on the framework of atomic norm minimization~\cite{bhaskar2013atomic}. In particular, Ref.~\onlinecite{bhaskar2013atomic} shows that, under a canonically chosen denoising parameter (which differs from the Fourier mode thresholding considered in our approach), the mean-squared reconstruction error scales asymptotically as $\epsilon\sqrt{\log(K)/K}$. Thus, we fix the threshold $\tau_r$ at the optimum identified in \cref{fig:lih_cr2_denoising_bound} and plot both bounds as functions of $K$. Numerically, our bound yields favorable reductions in the reconstruction error over the entire range of $K$ considered, supporting the performance gain of FDODMD over baseline ODMD shown in \cref{sec:results}.

\section{Formal Proof of \cref{theorem:recon}}
\label{app:formal_proofs}
To recap, the noiseless, noisy, and denoised data are defined by (\ref{eqn:doverlap}), (\ref{eqn:doverlapnoisy}), and (\ref{eqn:dift}), respectively;  $\mathbf{s}$, $\mathbf{d}$ and $\mathbf{r}$ denote their vectorized time domain representations.  Given a complex-valued vector $\bx = \{x_k\}_{k=0}^{K-1}$, we define the forward and inverse discrete Fourier transform (DFT) pair:
\begin{equation}
\label{eqn:dftpair}
\widehat{x}_k = \sum_{j=0}^{K-1} x_j\, e^{-2 \pi i \, j\, k/K} \ \text{ and } \
x_k = \frac{1}{K} \sum_{j=0}^{K-1} \widehat{x}_j\, e^{2 \pi i \, j\, k/K}.
\end{equation}
For any vector $\bx \in \mathbb{C}^K$, Plancherel's theorem states that
\begin{equation*}
    \| \bx \|_2^2 = \frac{1}{K} \| \widehat{\bx} \|_2^2,
\end{equation*}
where $ \| \bx \|_2 = (\bx^\dagger \bx)^{1/2} = \left( \sum_{k=0}^{K-1} | x_k |^2 \right)^{1/2}$ is the vector $2$-norm with $(\cdot)^\dagger$ denoting conjugate transpose (while we will use $(\cdot)^T$ for ordinary transpose). By Plancherel's theorem,
\begin{align}
\| \mathbf{r} - \mathbf{s} \|_2^2 &= \frac{1}{K} \| \widehat{\mathbf{r}} - \widehat{\mathbf{s}} \|_2^2 = \frac{1}{K} \sum_{k=0}^{K-1} | \widehat{r}_k - \widehat{s}_k |^2 \nonumber \\
 &= \frac{1}{K} \sum_{k=0}^{K-1} \bigl| \I_{ |\widehat{d}_k| \geq \tau_r } \widehat{d}_k - \widehat{s}_k \bigr|^2 \nonumber \\
 &= \frac{1}{K} \sum_{k=0}^{K-1} \bigl| \I_{ |\widehat{d}_k| \geq \tau_r } (\widehat{s}_k + \widehat{\xi}_k) - \widehat{s}_k \bigr|^2 \nonumber \\
\label{eqn:rminussprelim}
 &= \frac{1}{K} \sum_{k=0}^{K-1} \I_{ |\widehat{d}_k| \geq \tau_r } | \widehat{\xi}_k |^2 + \I_{ |\widehat{d}_k| < \tau_r } | \widehat{s}_k |^2.
\end{align}
To compute the expected value of the right-hand side, we must determine the distribution of $| \widehat{\xi}_k |^2$. The remainder of the proof proceeds by treating the noise in Fourier domain as a complex Gaussian process, and using properties of the exponential and error function distributions to compute the expected error.

To determine the distributions of the random vector $\bxi$ and its DFT $\widehat{\bxi}$, let $\MVN(\bmu, \bSigma)$ be the multivariate normal with PDF 
\begin{equation*}
f(\bw) = \frac{\exp \left(-{\frac{1}{2}}(\bw - \bmu)^T {\bSigma}^{-1}(\bw - \bmu)\right)}{(2\pi )^{K/2}\det({\bSigma})^{1/2}}.
\end{equation*}
Based on our noise assumptions in \cref{sect:noisemodels}, we have
\begin{equation}
\label{eqn:reimdistrib}
\Re \bxi \sim \MVN(\mathbf{0}, \bSigma) \ \text{ and } \ \Im \bxi \sim \MVN(\mathbf{0}, \bSigma),
\end{equation}
with $\boldsymbol{\Sigma} = \epsilon^2 \bI$. Let $\CN(\bmu, \bGamma, \bC)$ denote the complex normal with PDF
\begin{multline*}
    g(\bz) = \frac{1}{\pi^K\sqrt{\det(\bGamma)\det(\overline{\bGamma} - \bC^\dagger \bGamma^{-1} \bC )}}\, \\\cdot
    \exp \left\{-\frac12 \begin{bmatrix} \bz - \bmu \\ \overline{\bz} - \overline{\bmu}\end{bmatrix}^{\dagger}
    \begin{bmatrix}\bGamma & \bC \\ \overline{\bC}&\overline\bGamma\end{bmatrix}^{\!\!-1}\! \begin{bmatrix}\bz-\bmu \\ \overline{\bz}-\overline{\bmu}\end{bmatrix}
    \right\}.
\end{multline*}
Combining (\ref{eqn:reimdistrib}), statistical independence of $\Re \bxi$ and $\Im \bxi$, and properties relating the multivariate and complex normal distributions, we conclude that 
\begin{equation}
\label{eqn:xidistrib}
\bxi \sim \CN(\mathbf{0}, \bGamma, \mathbf{0}),
\end{equation}
with $\bGamma = 2 \bSigma = 2\epsilon^2 \bI$. Now we define the forward Fourier matrix $\mathcal{F}$ via
\begin{equation*}
\mathcal{F}_{jk} = e^{-2 \pi i j k/K},
\end{equation*}
for $0 \leq j, k \leq K-1$.  Note the use of zero-based indexing for row/column entries.  With this, the forward and inverse DFT (\ref{eqn:dftpair}) can be written as a matrix-vector multiplications,
\begin{equation}
\label{eqn:xihatmatvec}
\widehat{\bxi} = \mathcal{F} \bxi \quad \text{ and } \quad \bxi = \frac{1}{K} \mathcal{F}^\dagger \widehat{\bxi}.
\end{equation}
Hence $\mathcal{F}^\dagger \mathcal{F} = \mathcal{F} \mathcal{F}^\dagger = K \bI$. Modulo the normalization constant $K$, the matrix $\mathcal{F}$ is unitary.   Combining (\ref{eqn:xihatmatvec}) with (\ref{eqn:xidistrib}), we have
\begin{equation*}
\widehat{\bxi} \sim \CN(\mathbf{0}, \mathcal{F} \bGamma \mathcal{F}^\dagger, \mathbf{0}).
\end{equation*}
where $ \mathcal{F} \bGamma \mathcal{F}^\dagger = 2 \epsilon^2 \mathcal{F} \mathcal{F}^\dagger = 2 \epsilon^2 K \bI$. Hence
\begin{equation*}
\Re \widehat{\bxi} \sim \MVN(\mathbf{0}, \epsilon^2 K \bI) \ \text{ and } \
\Im \widehat{\bxi} \sim \MVN(\mathbf{0}, \epsilon^2 K \bI),
\end{equation*}
which means that each component is independent and has the same univariate normal distribution with mean $0$ and variance $\epsilon^2 K$. Standardizing, we have
\begin{equation*}
    \epsilon^{-1} K^{-1/2} \Re \widehat{\xi}_k, \epsilon^{-1} K^{-1/2} \Im \widehat{\xi}_k \sim \mathcal{N}(0, 1).
\end{equation*}
The sum of squares of two independent standard normals,
\begin{equation*}
    W = (\epsilon^{-1} K^{-1/2} \Re \widehat{\xi}_k)^2 + (\epsilon^{-1} K^{-1/2} \Im \widehat{\xi}_k)^2 = \epsilon^{-2} K^{-1} | \widehat{\xi}_k |^2,
\end{equation*}
has an exponential distribution with rate parameter $1/2$.  With this, we can conclude that $| \widehat{\xi}_k |^2$, which appears in the first summation of (\ref{eqn:rminussprelim}), is exponentially distributed with parameter $\lambda = \epsilon^{-2} K^{-1}/2$.  This implies that 
\begin{equation*}
    \E[ | \widehat{\xi}_k |^2 ] = \lambda^{-1} = 2 K \epsilon^2.
\end{equation*}
Summing this over $k$, we obtain $\E[ \| \widehat{\bxi} \|_2^2 ] = 2 K^2 \epsilon^2$:  Plancherel's theorem implies  $\E[ \| \bxi \|_2^2 ] = 2 K \epsilon^2$, which is indeed consistent with (\ref{eqn:xidistrib}).  We also conclude from the above that $| \widehat{\xi}_k |$ has a Rayleigh distribution with parameter $\sigma = 1/\sqrt{2 \lambda} = \epsilon K^{1/2}$.

On the other hand, we note that $| \widehat{d}_k | = | \widehat{s}_k + \widehat{\xi}_k | \leq | \widehat{s}_k| + |\widehat{\xi}_k |$, and $ \{ k \, | \, | \widehat{d}_k | \geq \tau_r \}  \subset  \{ k \, | \, | \widehat{s}_k| + |\widehat{\xi}_k | \geq \tau_r \}$.  Using this, we can take the expected value of the first term in (\ref{eqn:rminussprelim}) and derive
\begin{equation*}
    \E \Biggl[ \frac{1}{K} \sum_{k=0}^{K-1} \I_{ |\widehat{d}_k| \geq \tau_r } | \widehat{\xi}_k |^2 \Biggr] \leq   \frac{1}{K} \sum_{k=0}^{K-1} \E \Biggl[ \I_{ |\widehat{\xi}_k | \geq \tau_r - |\widehat{s}_k|   } | \widehat{\xi}_k |^2 \Biggr].
\end{equation*}
There are two cases.  When $\tau_r < |\widehat{s}_k|$, the indicator function on the right-hand side always remains $1$, so the expected value on the right-hand side is $2K \epsilon^2$ as derived above.  When $\tau_r \geq |\widehat{s}_k|$,
\begin{equation*}
    \E \Biggl[ \I_{ |\widehat{\xi}_k | \geq \tau_r - |\widehat{s}_k|   } | \widehat{\xi}_k |^2 \Biggr] = \E \Biggl[ \I_{ |\widehat{\xi}_k |^2 \geq (\tau_r - |\widehat{s}_k|)^2   } | \widehat{\xi}_k |^2 \Biggr].
\end{equation*}
As we know the distribution of $| \widehat{\xi}_k |^2$, we can directly evaluate the expectation: for $\alpha \geq 0$,
\begin{equation*}
   \E \Biggl[ \I_{ |\widehat{\xi}_k |^2 \geq \alpha } | \widehat{\xi}_k |^2 \biggr] = \int_{ \alpha }^{\infty} w \lambda e^{-\lambda w}  \, dw = \frac{e^{- \alpha \lambda} }{\lambda} (1 + \alpha \lambda). 
\end{equation*}
For convenience, we define the function $q(z) = e^{-z} (1+z),$
in terms of which we have
\begin{multline}
\label{eqn:bound1}
\E \Biggl[ \frac{1}{K} \sum_{k=0}^{K-1} \I_{ |\widehat{d}_k| \geq \tau_r } | \widehat{\xi}_k |^2 \Biggr] 
\leq \\ \hspace{1cm} 2 K \epsilon^2  \left[  \frac{1}{K} \sum_{k=0}^{K-1} \I_{\tau_r \geq | \widehat{s}_k |} q\left( \frac{(\tau_r- |\widehat{s}_k| )^2}{2  \epsilon^{2} K} \right) + \I_{\tau_r < | \widehat{s}_k |}  \right].
\end{multline}
We now return to the second half of (\ref{eqn:rminussprelim}). Taking expected values, we have
\begin{equation}
\label{eqn:expvalpart2}
\E \Biggl[ \frac{1}{K} \sum_{k=0}^{K-1}  \I_{ |\widehat{d}_k| < \tau_r } | \widehat{s}_k |^2  \Biggr] =  \frac{1}{K} \sum_{k=0}^{K-1}  | \widehat{s}_k |^2 \mathbb{P} \Bigl( \Bigl|\widehat{s}_k + \widehat{\xi}_k \Bigr| < \tau_r \Bigr).  
\end{equation}
Note that
\begin{equation*}
\mathbb{P} \Bigl( \Bigl|\widehat{s}_k + \widehat{\xi}_k \Bigr| < \tau_r \Bigr) 
= \int_{\Omega_k} \frac{1}{2 \pi \epsilon^2 K} \exp \left( -\frac{x^2 + y^2}{2 \epsilon^2 K} \right) \, dx \, dy,
\end{equation*}
where $\Omega_k = \{ (x, y) \in \mathbb{R}^2 \, | \, (\Re \widehat{s}_k + x)^2 + (\Im \widehat{s}_k + y)^2 < \tau_r^2 \}$
is the disk centered at $(-\Re \widehat{s}_k, -\Im \widehat{s}_k)$ with diameter $2 \tau_r$. Clearly, $\Omega_k \subset S_k$ for a square $S_k$
with the same center and side length $2 \tau_r$. Therefore,
\begin{multline}
 \mathbb{P} \Bigl( \Bigl|\widehat{s}_k + \widehat{\xi}_k \Bigr| < \tau_r \Bigr) \leq \int_{S_k} \frac{1}{2 \pi \epsilon^2 K} \exp \left( -\frac{x^2 + y^2}{2 \epsilon^2 K} \right) \, dx \, dy \\
  \leq \frac{1}{4} \left[ \Erf \left( \frac{ -\Re \widehat{s}_k + \tau_r} { \sqrt{2 \epsilon^2 K } } \right) - \Erf \left( \frac{ -\Re \widehat{s}_k - \tau_r} { \sqrt{2 \epsilon^2 K } } \right) \right]\\\cdot \left[ \Erf \left( \frac{ -\Im \widehat{s}_k + \tau_r} { \sqrt{2 \epsilon^2 K } } \right) - \Erf \left( \frac{ -\Im \widehat{s}_k - \tau_r} { \sqrt{2 \epsilon^2 K } } \right) \right],
\end{multline}
where $\Erf(z) = 2/{\sqrt{\pi}}\int_0^z e^{-t^2} \, dt$ denotes the error function.
Combining this with (\ref{eqn:expvalpart2}) and (\ref{eqn:bound1}), we can estimate the expected value of the RHS of (\ref{eqn:rminussprelim}), resulting in
\begin{multline}
\E \left[ \| \mathbf{r} - \mathbf{s} \|_2^2 \right] \\
\leq 
2 K \epsilon^2  \left[  \frac{1}{K} \sum_{k=0}^{K-1} \I_{\tau_r \geq | \widehat{s}_k |} q\left( \frac{(\tau_r- |\widehat{s}_k| )^2}{2  \epsilon^{2} K} \right) + \I_{\tau_r < | \widehat{s}_k |}  \right] \\
+  \sum_{k=0}^{K-1}  \frac{| \widehat{s}_k |^2 }{4K} \left[ \Erf \left( \frac{ -\Re \widehat{s}_k + \tau_r} { \sqrt{2 \epsilon^2 K } } \right) - \Erf \left( \frac{ -\Re \widehat{s}_k - \tau_r} { \sqrt{2 \epsilon^2 K } } \right) \right] \\
\cdot \left[ \Erf \left( \frac{ -\Im \widehat{s}_k + \tau_r} { \sqrt{2 \epsilon^2 K } } \right) - \Erf \left( \frac{ -\Im \widehat{s}_k - \tau_r} { \sqrt{2 \epsilon^2 K } } \right) \right],
\end{multline}
where we can identify the RHS with the function $G(\tau_r,K, \epsilon, \mathbf{s} )$ in \cref{theorem:recon}. $\Box$

\end{document}